\documentclass[transmag]{IEEEtran}
\usepackage{acronym}
\acrodef{3GPP}{3rd generation partnership project}
\acrodef{3D}{three-dimensional}
\acrodef{2D}{two-dimensional}
\acrodef{5G}{fifth generation}
 \acrodef{ABS}{almost blank subframe}
 \acrodef{AES}{advanced encryption standard}
\acrodef{APE}{absolute percentage error}
\acrodef{AR}{augmented reality}
    \acrodef{BS}{base station}
		\acrodef{BA}{bundle adjustment}
    \acrodef{CDF}{cumulative distribution function}
		\acrodef{CEVP}{constrained eigenvalue problem}
    \acrodef{CSI}{channel state information}
    \acrodef{CQI}{channel quality indicator}
    		\acrodef{CNN}{convolutional neural network}
		\acrodef{CSP}{communications service provider}
		\acrodef{CSI}{channel state information}
\acrodef{DL}{downlink}
\acrodef{DoF}{degree of freedom}
\acrodef{DNN}{deep neural network}
\acrodef{DQN}{deep Q-network}
\acrodef{DRL}{deep reinforcement learning}
\acrodef{DUDe}{downlink and uplink decoupling}
\acrodef{DDPG}{Deep Deterministic Policy Gradient}
\acrodef{DIRP}{distributed inter-cell inter-slice resource partition}

\acrodef{eICIC}{enhanced intercell interference coordination}
\acrodef{ESD}{energy spectral density}
\acrodef{ECDSA}{elliptic curve digital signature algorithm}
\acrodef{XR}{extended reality}
\acrodef{FDD}{frequency division duplex}
    \acrodef{FDMA}{frequency division multiple access}
		\acrodef{fps}{frame per second}
   \acrodef{GP}{Gaussian process}
    \acrodef{GPS}{global positioning system}
		\acrodef{GUI}{graphical user interface}
\acrodef{HetNet}{heterogeneous network}
\acrodef{HOG}{histogram of oriented gradients}
    \acrodef{ICI}{inter-cell interference}
		\acrodef{IMI}{inter-mode interference}
		\acrodef{IMU}{inertial measurement unit}
		\acrodef{IoT}{internet of things}
\acrodef{KL}{Kullback-Leibler}
\acrodef{KPI}{key performance indicator}
\acrodef{LTE}{long term evolution}

\acrodef{MAC}{medium access control}
\acrodef{mAP}{mean average precision}
\acrodef{M2M}{machine-to-machine}
\acrodef{MAP}{multi-agent process}
\acrodef{MAC}{media access control}
\acrodef{MAPE}{mean absolute percentage error}
\acrodef{MARL}{multi-agent reinforcement learning}
\acrodef{MADRL}{multi-agent deep reinforcement learning}
\acrodef{MDP}{Markov Decision Process}
\acrodef{MMDP}{multi-agent Markov Decision Process}
\acrodef{MIMO}{multiple-input and multiple-output}
\acrodef{MRU}{minimum resource unit}
\acrodef{mmWave}{millimeter wave}
\acrodef{MLP}{multi-layer perception}
\acrodef{MILP}{mixed integer linear programming}

\acrodef{NLES}{nonlinear equation system}
\acrodef{NSM}{Network Slicing Management}
   \acrodef{OFDM}{orthogonal frequency division multiplexing}
	\acrodef{ORB}{oriented FAST and rotated BRIEF}
\acrodef{OAM}{operations, administration, and maintenance}
    \acrodef{PDF}{probability density function}
    \acrodef{PPO}{proximal policy optimization}
    \acrodef{PHY}{physical layer}
		\acrodef{PSD}{power spectral density}
    \acrodef{PRB}{physical resource block}
	\acrodef{PRBs}{physical resource blocks}
   \acrodef{QoE}{quality of experience}
    \acrodef{QoS}{quality of service}
    \acrodef{RAN}{radio access network}
		\acrodef{RANSAC}{random sample consensus}
		\acrodef{RB}{resource block}
		\acrodef{RBS}{removal of bottleneck services}
		\acrodef{RL}{reinforcement learning}
		\acrodef{R-FCN}{region-based fully convolutional networks}
		\acrodef{RMDI}{resource muting for dominant interferer}
		\acrodef{RMSD}{root mean square distance}
		\acrodef{ROI}{region of interest}
		\acrodef{RPN}{region proposal network}
    \acrodef{RRM}{radio resource management}
		\acrodef{RU}{resource unit}
		\acrodef{RX}{receiver}
 \acrodef{SAFP}{successive approximation of fixed point}
 \acrodef{SE}{secure element}
    \acrodef{SDN}{software defined network}
    \acrodef{SNR}{signal-to-noise ratio}
    \acrodef{SINR}{signal-to-interference-plus-noise ratio}
\acrodef{SIR}{signal-to-interference ratio}
\acrodef{SIF}{standard interference function}
\acrodef{SLA}{service level agreement}
\acrodef{SLAM}{simultaneous localization and mapping}
\acrodef{SLAMORE}{Simultaneous Localization and Mapping with Object REcognition}
\acrodef{SoC}{system-on-chip}
\acrodef{SONs}{self organizing networks}
\acrodef{SSD}{single-shot multibox detector}
\acrodef{SPI}{serial peripheral interface}
    \acrodef{SVM}{support vector machine}
		\acrodef{SVD}{singular value decomposition}
    \acrodef{TCP}{transmission control protocol}
		\acrodef{TLS}{transport layer security}
		\acrodef{TL}{transfer learning}
		\acrodef{TDD}{time division duplex}
	\acrodef{TD3}{Twin Delayed Deep Deterministic policy gradient}
    \acrodef{TDMA}{time division multiple access}
		\acrodef{TTI}{transmission time interval}
		\acrodef{TX}{transmitter}
		\acrodef{UART}{universal asynchronous receiver-transmitter}
		\acrodef{UDP}{user datagram protocol}
		\acrodef{UE}{user equipment}
		\acrodef{UI}{user interface}
		\acrodef{UL}{uplink}
		\acrodef{UAV}{unmanned aerial vehicle}
		\acrodef{VO}{visual odometry}
		\acrodef{V2X}{vehicle-to-everything}
    \acrodef{WLAN}{wireless local area network}
\acrodef{YOLO}{you only look once}
\usepackage[left=1.57cm,right=1.57cm,top=0.95cm,bottom=2.54cm]{geometry}
\usepackage[font=small,skip=2pt]{caption}
\usepackage{cite}
\usepackage{amsmath,amssymb,amsfonts}
\usepackage[utf8]{inputenc}
\usepackage[mathscr]{euscript}
\usepackage{algorithm}
\usepackage{algpseudocode}
\algnewcommand{\Initialize}[1]{%
  \State \textbf{Initialize:}
  \Statex \hspace*{\algorithmicindent}\parbox[t]{.8\linewidth}{\raggedright #1}
}
\usepackage{graphicx}
\usepackage{textcomp}
\usepackage{xcolor}
\def\BibTeX{{\rm B\kern-.05em{\sc i\kern-.025em b}\kern-.08em
    T\kern-.1667em\lower.7ex\hbox{E}\kern-.125emX}}

\usepackage{url}
\usepackage{bbm}

\newtheorem{problem}{Problem}
\newtheorem{definition}{Definition}

\DeclareMathOperator*\argmax{arg \, max}		

\newcommand{\field}[1]{\mathbb{#1}}

\newcommand{\set}[1]{\mathcal{#1}}

\newcommand{\operator}[1]{\mathrm{#1}}
\newcommand{\RNum}[1]{\uppercase\expandafter{\romannumeral #1\relax}}

\newcommand{\R}{{\field{R}}}   
\newcommand{\Ex}{{\field{E}}}
\newcommand{\NN}{{\field{N}}}

\newcommand{\g}{\operator{(G)}}
\newcommand{\s}{\operator{(S)}}
\newcommand{\ev}{\operator{(Eval)}}
\newcommand{\ex}{\operator{(Expl)}}
\newcommand{\train}{\operator{(Train)}}

\newcommand{\ve}[1]{\boldsymbol{\mathbf{#1}}} 

\newcommand{\vs}{\ve{s}}

\newcommand{\vm}{\ve{m}}

\newcommand{\vc}{\ve{c}}
\newcommand{\va}{\ve{a}}

\newcommand{\N}{{\set{N}}}
\newcommand{\B}{{\set{B}}}
\newcommand{\K}{{\set{K}}}

\newcommand{\T}{{\set{T}}}
\newcommand{\M}{{\set{M}}}
\newcommand{\D}{{\set{D}}}
\newcommand{\Ss}{{\set{S}}}
\newcommand{\A}{{\set{A}}}

\newcommand{\X}{\set{X}}
\newcommand{\Y}{\set{Y}}
\newcommand{\Ho}{\set{H}}

\def\BibTeX{{\rm B\kern-.05em{\sc i\kern-.025em b}\kern-.08em T\kern-.1667em\lower.7ex\hbox{E}\kern-.125emX}}
\begin{document}

\title{Inter-Cell Network Slicing with Transfer Learning Empowered Multi-Agent Deep Reinforcement Learning}


\author{Tianlun Hu, \IEEEmembership{Student Member, IEEE}, Qi Liao, \IEEEmembership{Member, IEEE}, Qiang Liu, \IEEEmembership{Member, IEEE}, and Georg Carle
%
\thanks{Tianlun Hu is with both Nokia Bell Labs, Stuttgart, Germany, and Technical University of Munich, Munich, Germany (e-mail: tianlun.hu@nokia.com).}
\thanks{Qi Liao is with Nokia Bell Labs, Stuttgart, Germany (e-mail: qi.liao@nokia-bell-labs.com).}
\thanks{Qiang Liu is with University of Nebraska-Lincoln, Lincoln, USA (e-mail: qiang.liu@unl.edu).}
\thanks{Georg Carle is with Technical University of Munich, Munich, Germany (e-mail: carle@net.in.tum.de).}
\thanks{This work was supported by the German Federal Ministry of Education
and Research (BMBF) project KICK [16KIS1102K].}
\thanks{Partial contents of this paper appear in International Conference on Communications (ICC) 
 2022~\cite{Hu2022InterCellReDRL}.}
}

\IEEEtitleabstractindextext{\begin{abstract}
Network slicing enables operators to cost-efficiently support diverse applications on a common physical infrastructure.
The ever-increasing densification of network deployment leads to complex and non-trivial inter-cell interference, which requires more than inaccurate analytic models to dynamically optimize resource management for network slices. 
In this paper, we develop a DIRP algorithm with multiple deep reinforcement learning (DRL) agents to cooperatively optimize resource partition in individual cells to fulfill the requirements of each slice, based on two alternative reward functions with max-min fairness and logarithmic utility.
Nevertheless, existing DRL approaches usually tie the pretrained model parameters to specific network environments with poor transferability, which raises practical deployment concerns in large-scale mobile networks. 
Hence, we design a novel transfer learning-aided DIRP (TL-DIRP) algorithm to ease the transfer of DIRP agents across different network environments in terms of sample efficiency, model reproducibility, and algorithm scalability.
The TL-DIRP algorithm first centrally trains a generalized model and then transfers the \lq\lq generalist\rq\rq \ to each local agent (a.k.a., the \lq\lq specialist\rq\rq) with distributed finetuning and execution.
TL-DIRP consists of two steps: 1) centralized training of a generalized distributed model, and 2) transferring the \lq\lq generalist\rq\rq \ to each local agent with distributed finetuning and execution. 
We comprehensively investigate different types of transferable knowledge: model transfer, instance transfer, and combined model and instance transfer.
We evaluate the proposed algorithms in a system-level network simulator with $12$ cells.
The numerical results show that not only DIRP outperforms existing baseline approaches in terms of faster convergence and higher reward, but more importantly, TL-DIRP significantly improves the service performance, with reduced exploration cost, accelerated convergence rate, and enhanced model reproducibility. 
As compared to a traffic-aware baseline, TL-DIRP provides about $15\%$ less violation ratio of the \ac{QoS} for the worst slice service and $8.8\%$ less violation on the average service \ac{QoS}.

\end{abstract}

}

\maketitle

\section{INTRODUCTION}
\IEEEPARstart{E}{merging} technologies, e.g., autonomous driving, augmented and mixed reality, lead to increasingly volatile network dynamics in terms of traffic, mobility, and demand. To cost-efficiently accommodate heterogeneous services with diverse performance requirements, \acp{CSP} offer virtual end-to-end networks (a.k.a., slices) on common shared network physical infrastructures, e.g., base stations and network switches.
Network slicing enables performance and functional isolation, which guarantees that the slice performance is not affected by the operations in other slices, and assures the manageability for their slice tenants, respectively.
To achieve dynamic network slicing under varying slice traffic, efficient resource management of virtual network resource is necessitated. 
For instance, a variety of slice-aware scheduling algorithms \cite{ksentini2017toward,vo2018slicing} are proposed in \ac{RAN} to dynamically allocate radio resource (e.g., \acp{PRB}) of individual base stations, e.g., eNBs and gNBs, to different slices according to network conditions and service demands.

\begin{figure}[t]
	\centering
	\includegraphics[width=0.48\textwidth]{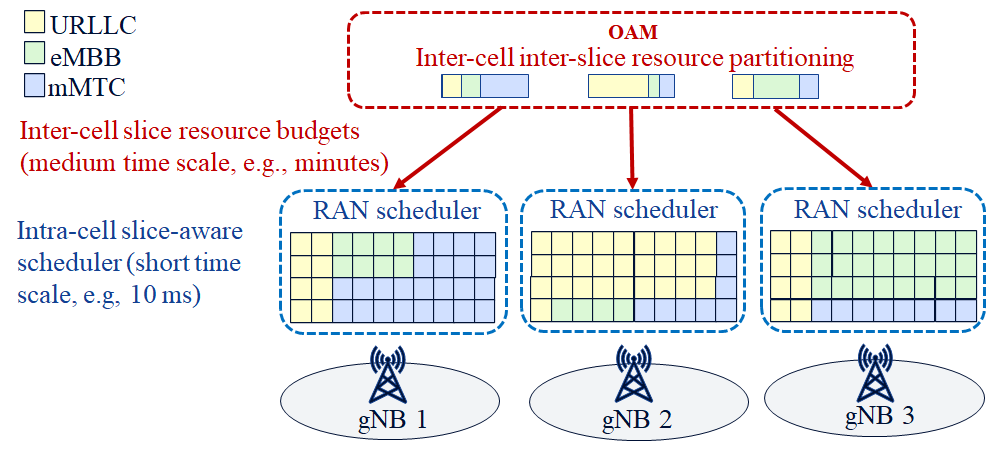}
	\caption{Dynamic multi-cell slicing resource allocation}
	\label{fig:RANSlicing}
\end{figure}

With the increasing spread of base station deployment in 5G and beyond, network slicing is becoming more complex.
As a result, the lack of interference coordination in existing individualized approaches can degrade the slice performance in multi-cell scenarios~\cite{Hu2022InterCellReDRL}.
Many works proposed model-based resource allocation and scheduling algorithms with inter-cell coordination, which rely on the approximated mathematical models towards the fast-changing interference and various optimization methods, e.g., linear programming \cite{Addad2020OptimizationMF, Beshley2021QoSAwareOR} and convex optimization \cite{Fossati2020MultiResourceAF,Ma2020SlicingRA}. These algorithms are proposed to be implemented in \ac{RAN}, and, to model the network capacity, they usually assume perfect \ac{CSI} shared among all cells. In practical systems, however, such algorithms are extremely difficult to implement, because of two reasons at least: first, \ac{RAN} scheduler makes decisions at a short time scale, e.g., every $10$ ms, while such time constraint is very challenging for the model-based algorithms due to the high communication overhead (caused by \ac{CSI} exchange) and the high computational cost (the \lq\lq snapshot\rq\rq \ approaches cannot well adapt to network dynamics and need to solve the problem for every time slot); second, these analytical solutions tend to fail in the real networks, because the approximated models cannot fully and accurately represent the complex network dynamics. Thus, in practical systems, as shown in Fig. \ref{fig:RANSlicing}, inter-cell inter-slice resource partitioning is introduced into network \ac{OAM} \cite{TS123501}, which collects a limited set of \acp{KPI} from all cells at medium time scale (e.g., minutes or even a quarter hour), performs inter-slice resource partitioning, and provides per-slice resource budgets to all cells. Then, each \ac{RAN} receives the resource budgets computed by \ac{OAM} periodically and uses them as resource constraints to guide the slice-aware scheduling and \ac{PRB} allocation algorithms within \ac{RAN}. In this paper, we focus on the inter-cell inter-slice resource partitioning problem in network \ac{OAM}. Note that in \ac{OAM} we tackle a problem different from the conventional coordinated interference mitigation problem at the \ac{MAC} layer in \ac{RAN}, because we obtain only a limited set of cell-based \acp{KPI}, while the short-term physical layer measurements such as \ac{CSI} are not available. Moreover, the mapping from the multi-cell network optimization parameters to these higher-layer \acp{KPI} is usually non-linear and non-convex, and the optimization goal is often multi-objective. Thus, \ac{OAM} usually benefits from the model-free machine learning and deep learning approaches that can be implemented in a distributed, cloud-native manner.

Recent advances in model-free approaches, especially \ac{DRL} \cite{Mao2016ResourceMW,ConstrainedRLNetSlicing}, have shown promising potential in automatically learning to manage radio access networks without the need for prior models.
In general, the resource management problem is formulated as a \ac{MDP}, which is then addressed by training and deriving a deep neural network parameterized policy.
A variety of \ac{DRL} algorithms, e.g., \ac{DQN}, \ac{DDPG}, and \ac{PPO} are exploited to achieve better policies in terms of performance, robustness, and convergence.
In particular, the problems with constraints, e.g., performance requirements, are resolved by leveraging different methods, e.g., interior-point policy optimization~\cite{ConstrainedRLNetSlicing} and Lagrangian primal-dual methods~\cite{DeepSlicingQiang}.
The inter-cell coordination problem is studied with distributed \ac{MADRL} approaches, which create multiple DRL agents and train their policies in different schemes.
The centralized scheme aims to train a common policy for all agents, where agents are distributedly executed with the shared model as the training completes.
For example, Li \emph{et al.} \cite{Li2018DeepRL} proposed a centralized scheme for slicing resource management with a DRL-based algorithm, but it fails to address the model complexity of agents when the network scale grows.
In contrast, the distributed scheme\cite{Alqerm2016ACO,Zhao2019DeepRL,Shao2021GraphAN} independently trains agents with individualized policy, which shows promising performance improvement in terms of convergence speed and communication overhead.
Zhao \emph{et al.} \cite{Zhao2019DeepRL} investigated the dynamic resource allocation problem in network slicing with distributed \ac{DRL}, which lacks inter-agent coordination and thus suffers uncoordinated interference in multi-cell slicing management.
Several efforts~\cite{Nie2021MultiAgentDR} have been made to address the issue of non-stationary environments from the perspective of individual agents, e.g., augmenting the state space of individual agents.
However, these aforementioned approaches raise concerns about sample efficiency, lengthy exploration, and convergence speed, which hinder their practical implementations in large-scale networks.

The emerging \ac{TL} techniques~\cite{pan2009survey} have been increasingly studied to address this challenge regarding the algorithm scalability, model reproducibility, and sample efficiency in machine learning-based approaches~\cite{nguyen2021transfer,wang2021transfer,Parera20}.
The basic idea of \ac{TL} is to utilize prior knowledge from pretrained models to benefit the learning process in target models.
Although there are extensive \ac{TL} works~\cite{Taylor2007TransferLV,zhuang2020comprehensive}, they are in the supervised learning domain, e.g., computer vision, and cannot be directly applied in \ac{RL} domain~\cite{Taylor2009TransferLF,zhu2020transfer}.
A few works\cite{nagib2021transfer, mai2021transfer} studied \ac{TL} in resource allocation in mobile networks, e.g., spectrum sharing in \ac{V2X}~\cite{Zafar2021TransferLI} and parameter optimization in network slicing.
However, \ac{TL}-assisted \ac{MADRL} in inter-cell network slicing scenarios is still an open problem.

In this paper, we focus on the inter-cell resource partition problem in network slicing with distributed \ac{MADRL} by extending our previous work~\cite{Hu2022InterCellReDRL}.
Our objective is to optimize the service qualities over all slices and cells while satisfying the constraints of the resource capacity.
We first develop a \ac{DIRP} algorithm, which effectively solves the problem with an inter-agent coordination mechanism, allowing information sharing between cells. The optimization is based on two alternative designs of objectives: 1) \emph{max-min fairness} over all slices, and 2) \emph{maximizing the average logarithmic utility} over all slices. The former guarantees that all slice-specific requirements for throughput and delay are fulfilled. The motivation is to align with 3GPP specifications that the service provided by any network slice must comply with the \ac{SLA} \cite{TS28530}. Note that max-min fairness, known to provide the best fairness guarantees, is a special case of the general class of the well-known $\alpha$-fair utility functions \cite{mo2000fair,bonald2006queueing}. The latter, as a classical concave utility function, also belonging to the $\alpha$-fair utility functions, compromises the \ac{SLA} fairness to improve resource efficiency \cite{AutonomicP}.
Then, we design a transfer learning-aided DIRP (TL-DIRP) algorithm to further improve the sample efficiency, model reproducibility, and algorithm scalability. We investigate the effectiveness of the transferable knowledge in three schemes, i.e., \emph{pretrained model transfer}, \emph{instance transfer}, and \emph{combined model and instance transfer}.
We further observe several key insights from the simulation results when integrating \ac{TL} in \ac{MADRL} under these schemes.
The contributions of this paper are summarized as follows:
\begin{itemize}
    \item We formulate the dynamic inter-cell resource partitioning problem to meet the requirements of throughput and latency for all slices, under the inter-slice resource constraints. We study two alternative objectives: 1) maximizing the minimum service quality over all slices and cells, and 2) maximizing the average of logarithmic utilities over all slices.
    \item We design a multi-agent \ac{DRL} algorithm to solve the problem with inter-agent coordination. We show that inter-agent load sharing improves the performance of conventional distributed schemes while achieving a lower model complexity and a faster convergence in comparison with centralized single-agent schemes.
    \item We further design a novel TL-DIRP algorithm to ease the transfer of DIRP agents across different network environments and analyze its effectiveness in three schemes, i.e., \emph{pretrained model transfer}, \emph{instance transfer} and \emph{combined model and instance transfer}. 
    \item We implement the proposed solutions in a system-level simulator and evaluate by comparing them with three baselines, i.e., centralized \ac{DRL}, distributed \ac{DRL}, and a traffic-aware heuristic approach. 
    The results show that \ac{DIRP} outperforms all three baselines in terms of per-slice service quality, and the proposed TL-DIRP further improves the performance with much faster algorithm convergence and lower exploration cost.
\end{itemize}

The rest of the paper is organized as follows. In Section \ref{sec:model}, we define the system model and formulate the inter-cell inter-slice resource partitioning problem. In Section \ref{sec:MADRL}, we propose the DIRP algorithm to solve the problem with inter-agent coordination. In Section \ref{sec:TL-MADRL}, we enhance the DIRP algorithm with transfer learning and investigate different types of transferable knowledge. The numerical results are demonstrated in Section \ref{sec:simu}. Finally, we conclude this paper in Section \ref{sec:concl}.

\section{RELATED WORK}
This work relates to network resource management, deep reinforcement learning in mobile networks, and transfer learning in networking.

\textbf{Model-based resource management.}
There are extensive works that use model-based approaches to manage the resource allocation of \ac{RAN} slices in 5G and beyond networks. 
Several works \cite{Fossati2020MultiResourceAF, Ma2020SlicingRA} investigated the problem of network slice resource allocation by assuming the resource demands are known and static and leveraged the methods of convex optimization to solve the problem with different utility functions.
The network slicing for machine-type communications is studied in \cite{Beshley2021QoSAwareOR}, where a radio resource allocation method is proposed to dynamically select channel bandwidth according to the \ac{QoS} requirements and traffic aggregation in \ac{M2M} gateways.
Addad \emph{et al.} \cite{Addad2020OptimizationMF} analyzed the virtual network function deployment in network slicing, formulated a mixed-integer linear programming model, and proposed a heuristic algorithm under different resource constraints.
Cavalcante \emph{et al..} \cite{Cavalcante2019ConnectionsBS} formulated a max-min fairness problem to handle load-coupled interference, then transformed it into a fixed point problem and solved it with low complexity iteration algorithm. 
Recently, an inter-cell coordinated scheme for dense cellular network resource scheduling was proposed \cite{Sciancalepore2018AMI}, which tackled inter-cell interference and provided inspiring results.
However, the approximated mathematical models cannot fully represent the characteristics of complex networks. More importantly, applying these model-based solutions in \ac{OAM} is challenging due to the lack of \ac{CSI} measurements at fine time granularity.

\textbf{Deep reinforcement learning in mobile networks.}
Liu \textit{et al..} \cite{ConstrainedRLNetSlicing} proposed a constrained \ac{DRL}-based on interior-point policy optimization (IPO) to solve the slicing resource allocation problem in the single base station scenario.
Xu \emph{et al..} \cite{Xu2018ExperiencedrivenNA}, studied a \ac{DRL}-based solution to extract per-slice users' behavior with traffic-aware exploration and allocate sufficient \ac{RAN} resource accordingly.
Liu \emph{et al..} \cite{DeepSlicingQiang} proposed a \ac{DRL}-based algorithm named DeepSlicing by decomposing \ac{RAN} slicing optimization into a master problem and several slave problems, which are addressed with a joint coordinator and associated \ac{DRL} agent for each slice respectively.
However, these works are designed to address the resource allocation problem in single-cell scenarios. 
Several works \cite{Alqerm2016ACO, Zhao2019DeepRL} studied the multi-cell scenarios and proposed several \ac{DRL} solutions with discrete action space.
Recent efforts \cite{Song2021ADR, Peng2020DeepRL} extended the discrete action space into continuous action space, which showed improved performances in handling complex scenarios.
However, none of them addressed the inter-cell dependencies and inter-slice resource constraints. 

\textbf{Transfer learning in networking.}
Xu \emph{et al..} proposed an aggregation \ac{TL} method applied to \ac{MADRL} for real-time strategy games by transferring knowledge from small-scale to large-scale multi-agent systems, which improves the convergence speed of the algorithm~\cite{Xu2021AggregationTL}. 
Zafar \emph{et al..} proposed to enhance the double Q-learning with \ac{TL} for solving the decentralized spectrum sharing problem in the \ac{V2X} communication networks~\cite{Zafar2021TransferLI}. 
By transferring the Q-values of the expert model to the learner model, the \ac{TL}-assisted method accelerates the convergence rate of the learner model.
Mai \emph{et al..} \cite{mai2021transfer} proposed to optimize the slice parameters, e.g., transmission power and spreading factor, with \ac{DDPG} and \ac{TL}. 
The \ac{TL} was conducted by pretraining a model on a centralized controller and then using it as the initial model on local slice optimization tasks. 
Nagib \emph{et al..} \cite{nagib2021transfer} studied \ac{TL} to accelerate the \ac{DRL} algorithms for dynamic \ac{RAN} slicing resource allocation in single-cell scenarios, by transferring the model pretrained from an expert base station to a learner base station.
Nevertheless, none of the above-mentioned works studied \ac{TL} in coordinated \ac{MADRL} for inter-cell slicing resource partition. 
\section{System Model and Problem Formulation}\label{sec:model}
In this section, we first describe the \ac{MDP}-based system model in Section \ref{subsec:sysmodel}. Then, we formulate the optimization problem based on the \ac{MDP} model in Section \ref{subsec:problemform}. Table \ref{Table_notation} summarizes the notations used in this work.


\begin{table}[ht]
\caption{Table of Notations}
\centering
\label{Table_notation}
\begin{tabular}{|l|l|}
\hline
Symbol & Meaning \\ \hline
   $\vs$    &     Global state in $\Ss$    \\\hline
   $\va$  &     Global action in $\A$    \\\hline
   $r$  &     Global reward \\\hline
   $\vs_k$   &     Local state in $\Ss_k$ of Agent $k$     \\\hline
   $\va_k$   &     Local action in $\A_k$ of Agent $k$     \\\hline
   $r_k$   &     Local reward of agent $k$  \\\hline
   $\tilde{r}_k$   &     Approximated local reward of Agent $k$  \\\hline
   $\vm_k$   &   Message sent from Agent $k$ to neighbors    \\\hline
   $\overline{\vm}_k$   &   Received messages from all neighbors of Agent $k$  \\\hline
   $\vc_k$   &    Extracted information from $\overline{\vm}_k$ of Agent $k$ \\\hline
   $Q_{\theta}$   &   Current critic network with parameter $\theta$  \\\hline
   $\pi_{\phi}$   &   Current actor network with parameter $\phi$  \\\hline
   $Q_{\theta'}$   &     Target critic network with parameter $\theta'$   \\\hline
   $\pi_{\phi'}$   &    Target actor network with parameter $\phi'$     \\\hline
   $\pi^{\g}$   &    Generalist's policy learned by central controller   \\\hline
   $\pi_k$   &     Specialist's policy learned by Agent $k$    \\\hline
   $\D_S$   &    Source domain  $\D_S:=\D^{\g}$, i.e., generalist's domain    \\\hline
   $\T_S$   &    Source task  $\T_S:=\T^{\g}$, i.e., generalist's task    \\\hline
   $\D_T$   &    Target domain $\D_T:=\D_k^\s, k\in\K$, i.e., specialist's domain  \\\hline
   $\T_T$   &    Target task $\T_T:=\T_k^\s, k\in\K$, i.e., specialist's task    \\\hline
   
\end{tabular}
\end{table}
\subsection{System Model}\label{subsec:sysmodel}
We consider a network system consisting of a set of cells $\K:=\left\{1, 2, \ldots, K\right\}$ and a set of slices $\N:=\left\{1, 2, \ldots, N\right\}$. Each slice $n\in\N$ has predefined throughput and delay requirements, denoted by $\phi_{n}^\ast$ and $d_{n}^\ast$, respectively. The network system runs on discrete time slots $t\in\NN_0$. \ac{OAM} adapts the inter-slice resource partitioning for all cells periodically to meet their performance requirements, as illustrated in Fig. \ref{fig:RANSlicing}.

To capture the temporal and inter-cell dependencies, we model the multi-cell resource partition as an \ac{MDP} defined by $\M:=\{\Ss, \A, P(\cdot), r(\cdot), \gamma \}$, where $P: \Ss\times\A\times\Ss\to [0, 1]$ denotes the transition probability distribution over state space $\Ss$ and action space $\A$. $r: \Ss\times\A\to\R$ is the reward function, which evaluates the per-slice \ac{QoS} for all cells and $\gamma\in[0, 1]$ denotes the discount factor for cumulative reward calculation.

Assuming that at each time step $t$, the network observes the global {\bf state} $\vs(t):=[\vs_1(t),\ldots,\vs_K(t)]\in\Ss$, where $\vs_k(t)$ is the local state observed from cell $k$. The {\bf action} at slot $t$ denoted by $\va(t):=[\va_1(t), \ldots, \va_K(t)]\in\A$, includes the \ac{RAN} slice resource budget, where the local action $\va_k(t)\in\A_k$ indicates the partitioning ratio $a_{k, n}(t)\in [0, 1]$ to each slice for $n\in\N$ aligning with intra-cell resource constraints. Thus, the local action space $\A_k$ and the global action space $\A$ yield
\begin{align}
	\A_k &:= \left\{\va_k\bigg|a_{k,n}\in[0,1], \forall n\in\N; \sum_{n=1}^{N} a_{k,n} = 1\right\}. \label{eqn:local_actionspace} \\
	\A & := \left\{\va\big|\va_k\in\A_k, \forall k\in \K\right\}. \label{eqn:global_actionspace}
\end{align}

The goal is to maximize the satisfaction level of \ac{QoS} in terms of throughput and delay requirements $(\phi_{n}^\ast, d_{n}^\ast)$ for every slice $n\in\N$ in each cell $k\in\K$.
Thus, we design two alternative {\bf reward} functions for the two alternative objective designs: \emph{max-min fairness} and \emph{maximizing the average logarithmic utilities}. The former provides the best fairness that guarantees overall slice requirements by giving the maximum protection to the most critical and resource-demanding slice. While the latter, although taking fairness into account, still tries to achieve a good fairness-efficiency tradeoff.

The global \textbf{reward} function $r(t)$, based on the two alternative objectives, respectively, is defined as follows:

\begin{enumerate}
    \item \emph{Max-min fairness}: we define $r(t)$ as the minimum per-slice \ac{QoS} satisfaction level based on the observed average throughput $\phi_{k,n}(t)$ and average delay $d_{k,n}(t)$ at time step $t$ for each slice $n$ in cell $k$, as
    \begin{equation}
	r(t) := \min_{k\in\K, n\in\N} \min\left\{\frac{\phi_{k,n}(t)}{\phi_{n}^\ast}, \frac{d_{n}^\ast}{d_{k,n}(t)}, 1\right\}.
	\label{eqn:global_rewardfunction_maxmin}
    \end{equation}
    The reward formulation drops below $1$ when the actual average throughput or delay of any slices fails to fulfill the requirements. Note that the reward is upper bounded by $1$ even if all slices achieve better performances than the requirements, to achieve more efficient resource utilization. The second item in \eqref{eqn:global_rewardfunction_maxmin} is {\bf inversely proportional} to the actual delay, namely, if the delay is longer than required, this term is lower than $1$.
    \item \emph{Maximizing the average logarithmic utilities}: we define $r(t)$ as the average logarithmic utilities over the service satisfaction levels of all slices, given by 
    \begin{equation}
        \begin{aligned}
            r(t) & := \frac{1}{K\cdot N} \cdot \\ & \sum_{k\in\K, n\in\N} \log \left(\min\left\{\frac{\phi_{k,n}(t)}{\phi_{n}^\ast}, \frac{d_{n}^\ast}{d_{k,n}(t)}\right\} + 1\right)
            \label{eqn:global_rewardfunction_log}
        \end{aligned}
    \end{equation}
     where the \emph{service satisfaction level} per slice per cell $\min\left\{\frac{\phi_{k,n}(t)}{\phi_{n}^\ast}, \frac{d_{n}^\ast}{d_{k,n}(t)}\right\}\geq 0$ is defined as the minimum between the throughput and delay satisfaction levels. Thus, if either throughput or delay does not meet the requirement, this term is below $1$. By adding an offset $1$ within the log function with base $2$, the per-slice logarithmic utility function is always non-negative. Note that unlike  \eqref{eqn:global_rewardfunction_maxmin}, the reward in \eqref{eqn:global_rewardfunction_log} is not upper bounded by $1$, because the service satisfaction level is not upper bounded. However, if all slices' requirements are exactly met, then we have $r(t)=1$.
\end{enumerate}

\subsection{Problem Formulation}\label{subsec:problemform}
The problem is to find the optimal policy $\pi:\Ss\to\A$, which decides the inter-cell inter-slice resource partitioning $\va\in\A$ based on the observation of network state $\vs\in\Ss$, to maximize the expectation of the cumulative discounted reward defined in Eq.~\eqref{eqn:global_rewardfunction_maxmin} or Eq.~\eqref{eqn:global_rewardfunction_log} of a trajectory for a finite time horizon $T$. The problem is given by: 
\begin{problem}
    \label{prob:RL}
    \begin{equation}
      \max_{\pi} \  \Ex_{\pi} \left[\sum_{t=0}^{T} \gamma^t r\big(\vs(t), \va(t) \big)\right], 
      \mbox{s.t. } \va\in\A,
      \label{eqn:problem}
    \end{equation}
    where $\A$ is defined by Eq.~\eqref{eqn:local_actionspace} and Eq.~\eqref{eqn:global_actionspace}, $r$ is given by Eq.~\eqref{eqn:global_rewardfunction_maxmin} or Eq.~\eqref{eqn:global_rewardfunction_log}.
\end{problem}

The challenges of solving the aforementioned problem are two-fold. First, the global reward functions depend on high-dimensional state and action spaces, and involve complex inter-cell dependencies, which are difficult to be accurately obtained in practical network systems. For example, increasing resource partition in one slice $n$ and cell $k$ improves its own service performance, however, it decreases the available resource allocated to other slices in the same cell and may aggravate the interference received in neighboring cells. 
Besides, because we aim at solving the inter-cell inter-slice resource partitioning problem in \ac{OAM}, only a limited set of \acp{KPI} (e.g., averaged cell throughput and delay) at a medium time scale (e.g., every $15$ minutes) is available. It is extremely difficult to derive closed-form expressions for the multi-cell network with the extracted data at the higher layers (above \ac{MAC} layer) of the network system.
Second, the dynamic of network systems, e.g., additional cell deployments, changes the properties of the problem, e.g., leading to expanded state and action space. This requires the solution of this problem to be efficient and scalable in terms of fast convergence speed, high sample efficiency, and low computational efforts.

\section{Distributed Inter-Cell Resource Partition}\label{sec:MADRL}
In this section, we propose the distributed inter-cell inter-slice resource partition (DIRP) algorithm based on the \ac{MADRL} approach with an inter-agent coordination scheme. Then, we briefly introduce the actor-critic method to solve the \ac{DRL} problem. Next, we propose the method to tackle the intra-cell resource constraint with modified \ac{DRL} network architecture.
\subsection{Proposed DIRP Algorithm}\label{ssec:DIRP}
In this part, we propose the DIRP algorithm with inter-agent coordination, which allows each agent to learn an individualized policy and make its own decision on the local action, based on local observations and neighboring information. In contrast to conventional centralized \ac{DRL} \cite{Li2018DeepRL}, which collects global observation from all slices and cells of the network system, the DIRP algorithm may not achieve the global performance as good as the centralized one due to the limited observation on the entire network. However, it may converge much faster and be more sample efficient by using a less complex model based on lower dimensional state and action spaces, and the coordination mechanism could improve the performance of distributed agents with additional side information about the environment.

To capture local network observations, each agent $k$ observes its {\bf local state} $\vs_k$. In particular, we include the following measurements and performance metrics:
\begin{itemize}
  \item Average per-slice user throughput $\left\{\phi_{k,n}: n\in\N\right\}$;
  \item Per-slice load $\left\{l_{k,n}: n\in\N\right\}$;
  \item Per-slice number of active users $\left\{u_{k,n}: n\in\N\right\}$;
  \item Per-slice throughput requirement $\{\phi_{k,n}^*: n\in\N\}$;
  \item Per-slice delay requirement $\{d_{k,n}^*: n\in\N\}$.
\end{itemize}

In conventional distributed \ac{DRL} approach, each agent $k$ in the $k$-th cell computes a {\bf local reward} $r_k$, and makes decision on the {\bf local action} $\va_k\in\A_k\subset [0, 1]^N$. The local reward for max-min fairness or maximizing average logarithmic utilities, based on the local observations, is given by
\begin{enumerate}
    \item \emph{Max-min fairness}: 
    \begin{equation}
	\label{eqn:local_rewardfunction_maxmin}
	r_k(t) := \min_{n\in\N} \min\left\{\frac{\phi_{k, n}(t)}{ \phi_{n}^\ast}, \frac{d_n^\ast}{d_{k,n}(t)}, 1\right\},
    \end{equation}
    \item \emph{Maximizing average logarithmic utilities}:
    \begin{equation}
         \begin{aligned}
             r_k(t) &:= \frac{1}{N} \cdot \\
             &\sum_{n\in\N}\log\left(\min\left\{\frac{\phi_{k,n}(t)}{\phi_{n}^\ast}, \frac{d_{n}^\ast}{d_{k,n}(t)}\right\} + 1\right).
         \end{aligned}
         \label{eqn:local_rewardfunction_log}
    \end{equation}
\end{enumerate}

Note that $r_k$ not only depends on the local state-action pair but also on the states and actions of other agents. The global reward yields $r(t) = \min_{k\in\K} r_k(\va(t), \vs(t))$ with local reward \eqref{eqn:local_rewardfunction_maxmin} or $r(t) = \frac{1}{K} \sum_{k\in\K} r_k(\va(t), \vs(t))$ with local reward \eqref{eqn:local_rewardfunction_log}.
We can approximate $r_k(\vs, \va)$ based on the local observations $(\vs_k, \va_k)$, denoted by $\tilde{r}_k(\vs_k(t), \va_k(t))$. However, the estimation can be inaccurate because it neglects the inter-cell dependencies and estimates local reward independently.

Thus, to capture the inter-agent dependencies, in \ac{DIRP} algorithm we let the agents communicate and exchange additional information with neighboring cells. Let each agent $k$ send a message $\vm_k$ to a set of its neighboring agents, denoted by $\K_k$. Then, each agent $k$ holds the following information: local state and action pair $(\vs_k, \va_k)$ and received messages $\overline{\vm}_k:=\left[\vm_i: i\in\K_k\right]$.
One option is to directly use all received messages $\overline{\vm}_k$ along with $(\vs_k, \va_k)$ to estimate $r_k(\vs, \va)$ with $\tilde{r}_k(\vs_k, \overline{\vm}_k, \va_k)$. However, if the dimension of the exchanged message is high, this increases the complexity of the local model.

An alternative is to extract the useful information $\vc_k\in\R^{Z^{(c)}}$ from the received messages $\overline{\vm}_k\in\R^{Z^{(m)}}$ with $g:\R^{Z^{(m)}}\to\R^{Z^{(c)}}:\overline{\vm}_k\mapsto\vc_k$, such that $Z^{(c)} \ll Z^{(m)}$, where $Z^{(m)}$ and $Z^{(c)}$ stand for the corresponding dimensions. 
We can then use $\tilde{r}_k(\vs_k, \vc_k, \va_k)$ to approximate $r_k$, by capturing the hidden information in the global state, while remaining the low model complexity.
Pioneer works such as \cite{Foerster2016LearningTC} proposed to learn the extraction of the communication messages by jointly optimizing the communication action with the reinforcement learning model. However, the joint training of multiple interacting models usually leads to extended convergence time and even diverged training. To provide a robust and efficient practical solution, we leverage domain knowledge to extract the information. Knowing that the inter-agent dependencies are mainly caused by the load-coupling inter-cell interference, we propose to let each agent $k$ communicate with its neighboring agents the slice-specific load information $l_{k,n}$, $\forall n\in\N$. Then, based on the exchanged load information, we compute the average per-slice neighboring load as the extracted information $\vc_k(t)$. Namely, we define a deterministic function
\begin{equation}
  \label{neighbor_info}
  \begin{aligned}
  g_k: & \R^{N|\K_k|}\to\R^N : [l_{i, n}:n\in\N, i\in\K_k] \mapsto \vc_k(t)\\
    \mbox{with } &  \vc_k(t):=\left[\frac{1}{|\K_k|}\sum_{i\in \K_k}l_{i,n}(t): n\in\N \right].
  \end{aligned}
\end{equation}

In this way, the \ac{DIRP} algorithm solves Problem \ref{prob:RL} with approximated local reward while considering the inter-cell dependencies by including neighboring information. 
Thus, the \ac{DIRP} algorithm approximates $r_k(\vs, \va)$ with $\tilde{r}_k(\vs_k, \vc_k, \va_k)$, decomposes Problem \ref{prob:RL} with $K$ independent subproblems, and finds the following local policies $\pi_k:\Ss_k\times\R^N\to\A_k$ for each \ac{DIRP} agent $k\in\K$:
\begin{equation}
    \label{eqn:local_comm_policy}
    \begin{aligned}
        \pi_k^{\ast} &= \argmax_{\pi_k; \va_k\in\A_k} \Ex_{\pi_k} \left[\sum_{t=0}^{T} \gamma_k^t \tilde{r}_k\big(\vs_k(t), \vc_k(t), \va_k(t) \big)\right].
    \end{aligned}
\end{equation}

\subsection{The Training of Agents}\label{ssec:AC-method}
In this part, we follow the actor-critic method \cite{Konda1999ActorCriticA} to train the agents, which has proven effective when dealing with high dimensional and continuous state space. Such method solves the optimization problem by using critic function $Q(\vs_t, \va_t|\theta)$ (in this subsection, we denote $\vs(t)$ and $\va(t)$ by $\vs_t$ and $\va_t$ respectively for brevity) to approximate the value function, i.e., $Q(\vs_t, \va_t|\theta) \approx Q^\pi(\vs_t, \va_t)$, and actor $\pi(\vs_t|\phi)$ to update the policy $\pi$ at every \ac{DRL} step in the direction suggested by critic. For brevity, we denote the network with parameters in the form $Q_\theta$ and $\pi_\phi$ for critic and actor respectively.

\begin{figure}[t]
	\centering
	\includegraphics[width=0.45\textwidth]{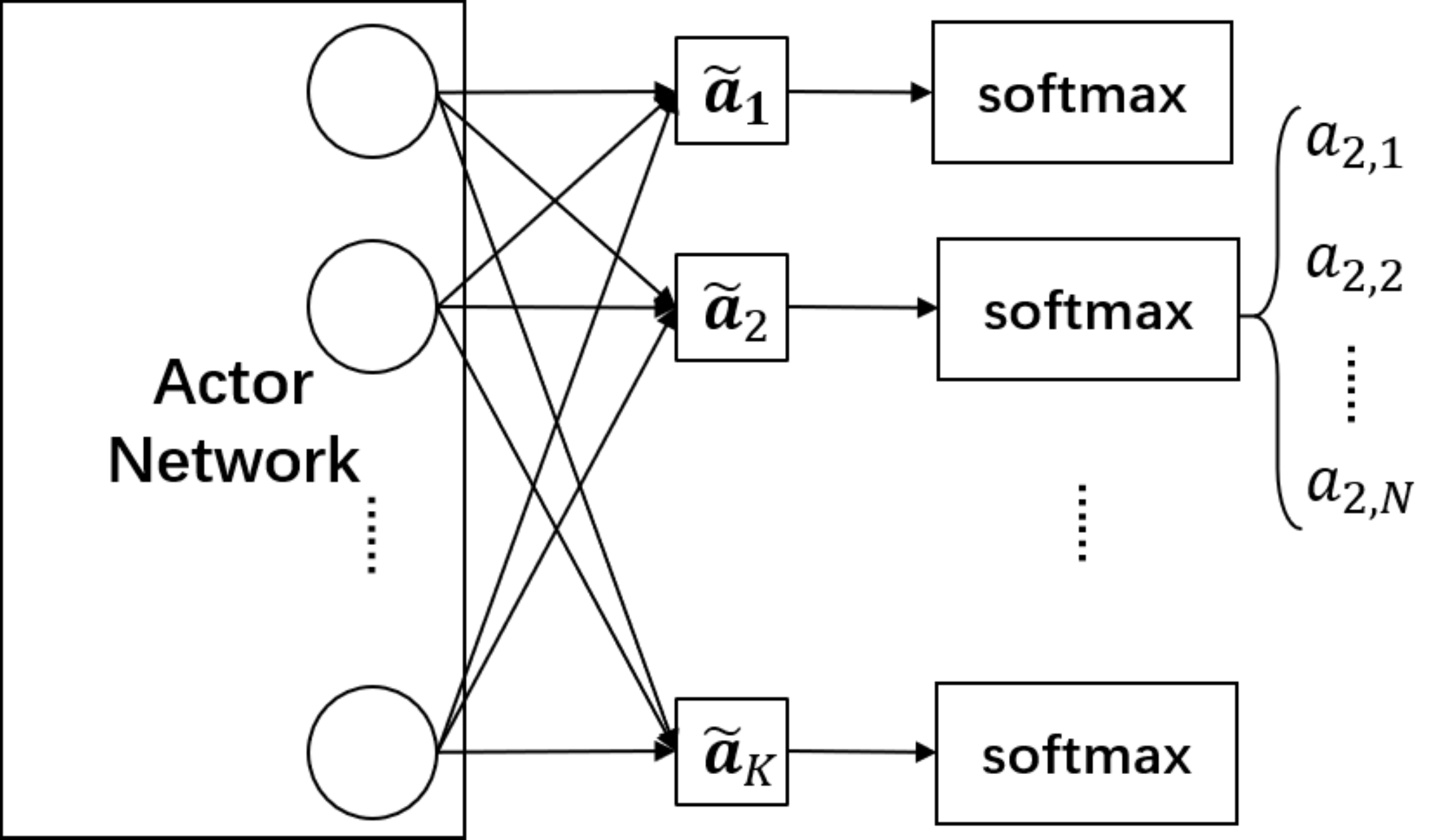}
	\caption{Actor's output layer with decoupled softmax activation}
	\label{fig:SepOutAct}
\end{figure}

In this work, we use \ac{TD3} algorithm \cite{Fujimoto2018AddressingFA} as an off-policy \ac{DRL} algorithm built on top of the actor-critic method. As an extension of \ac{DDPG} \cite{Silver2014DeterministicPG}, \ac{TD3} overcomes the \ac{DDPG}’s problem of overestimating Q-values by introducing a double critic structure for both current networks $Q_{\theta_1}, Q_{\theta_2}$ and target networks $Q_{\theta'_1}, Q_{\theta'_2}$. The minimum of the two Q-values is used to represent the approximated Q-value of the next state. 
Besides, the updates of the policy network are less frequent than the value network, which allows the value network to reduce errors before it is used to update the policy network. Moreover, \ac{TD3} uses target policy smoothing, i.e., adding noise to the target action, to make it harder for the policy to exploit Q-function errors by smoothing out Q along with changes in action.
The target actions are computed based on the next state collected in the sample, given by
\begin{equation}
    \label{eqn:target_action}
    \va'(\vs_{t+1}) = clip\left(\pi'_{\phi'}\left(\vs_{t+1} \right) + clip(\epsilon, -c, c), \va_{\operator{L}}, \va_{\operator{H}}\right)
\end{equation}
where the added noise $\epsilon\sim \N(0, \sigma)$ is clipped to keep the target close to the original action, and $\va_{\operator{L}}, \va_{\operator{H}}$ are the lower and upper bounds of the action, respectively. 

The target update in \ac{TD3} is given by:
\begin{equation}
		y_t = r_t + \gamma \min_{i=1,2} Q_{\theta'_i}\left(\vs_{t+1}, \va'(\vs_{t+1})\right).
		\label{eqn:TD3_target}
\end{equation}

The critic parameters $\theta_i, i\in\{1, 2\}$ are updated with temporal difference (TD) learning, given by:
\begin{equation}
\label{eqn:update-critic}
	L\left(\theta_i\right) = \Ex\left[\left(y_t - Q_{\theta_i}\left(\vs_t, \va_t\right)\right)^2\right].
\end{equation}

The actor is updated by policy gradient based on the expected cumulative reward $J$ with respect to the actor parameter $\theta^\pi$ with:
\begin{equation}
	\begin{aligned}
		\nabla_{\phi}J &\approx \Ex\left[\nabla_{\phi}Q_{\theta_1}\left(\vs, \va\right)|_{\vs=\vs_t, \va=\pi_{\phi}(\vs_t)}\right]\\
		&= \Ex\left[\nabla_{\va}Q_{\theta_1}\left(\vs, \va\right)|_{\vs=\vs_t, \va=\pi_{\phi}(\vs_t)}\nabla_{\phi}\pi_{\phi}\left(\vs_t\right)\right].
	\end{aligned}
	\label{eqn:TD3_actor_update}
\end{equation}

The parameters of the target networks are updated with the soft update to ensure that the TD-error remains small:
\begin{equation}
	\begin{aligned}
		&\theta'_i \leftarrow \tau\theta_i + (1-\tau)\theta'_i, i=1,2; \\
        &\phi' \leftarrow \tau\phi + (1-\tau)\phi'.
	\end{aligned}
	\label{eqn:Soft_update}
\end{equation}

\subsection{Dealing with Resource Constraints}\label{ssec:Softmax}
\label{subsec:constrant}
To address the inter-slice resource constraints in Eq.~\eqref{eqn:local_actionspace}, we propose a method by reconstructing the network architecture of \ac{DRL} model with an additional regularization layer. 

In this method, we embed a decoupled regularization layer into the output layer of the actor network, such that this layer becomes part of the end-to-end back propagation training of the neural network. Since the softmax function realizes for each $\va_k$ the following projection
$$
\sigma:\R^{N}\to\left\{\va_k\in\R^{N}\Big| a_{k,n}\geq 0, \sum_{n=1}^{N} a_{k,n} = 1\right\},
$$
the decoupled softmax layer well addresses the intra-cell inter-slice resource constraints $\sum_{n=1}^{N} a_{k,n} = 1$, $\forall k\in\K$ as shown in Fig. \ref{fig:SepOutAct}.


In summary, we provide the TD3-based DIRP algorithm with inter-agent coordination in Algorithm \ref{algo:TD3-DIRP}.
\begin{algorithm}
\caption{The DIRP Algorithm}
\label{algo:TD3-DIRP}
\begin{algorithmic}[1]
\State Initialize parameters for critics $Q_{\theta_1^k}$, $Q_{\theta_2^k}$ and actor $\pi_{\phi^k}$, with random parameters $\theta_1^k$, $\theta_2^k$, $\phi^k$, $\forall k\in\K$
\State Initialize target networks $\theta_1'^{k}\leftarrow\theta_1^k$, $\theta_2'^k\leftarrow\theta_2^k$, $\phi'^k\leftarrow\phi^k$
\State Initialize empty replay buffer $\B_k$
\State Initialize $\epsilon\in[0,1]$ and decay $d\in[0,1]$ for $\epsilon$-greedy exploration
\State Define time periods $\Ho^{\ex}, \Ho^{\train}, \Ho^{\ev}$ for exploration, training, and evaluation phases, respectively
\State {\bf Repeat}
\For{local agent $k\in K$}
\State Observe local state $\vs_k(t)$ and information $\vc_k(t)$
\State Select and execute action:
\If{$t\in \Ho^{\ex}$}
\State $\va_k(t) \leftarrow $ \emph{random choice}
\ElsIf{$t\in \Ho^{\train}$}
\State 
$\va_k(t)\leftarrow\left\{
\begin{aligned}
&\pi_k(\vs_k(t), \vc_k(t)) + \varepsilon , & \mbox{if } U[0,1] > \epsilon  \\
&\text{\emph{random choice}} , & \text{otherwise}
\end{aligned}\right.$
\State where $U[0,1]$ is the generated random value \State following uniform distribution in $[0,1]$.
\State $\epsilon\leftarrow d\epsilon$
\ElsIf{$t\in \Ho^{\ev}$}
\State $\va_k(t) = \pi_k(\vs_k(t), \vc_k(t))$
\EndIf
\State Observe next state $\vs_k(t+1)$, received information 
\State $\vc_k(t+1)$, and compute $r_k(t)$
\State Store instance in $\B_k$:
\State $\Big(\big(\vs_k(t),\vc_k(t)\big), \va(t), \big(\vs_k(t+1), \vc_k(t+1)\big), r_k(t)\Big)$
\If{time to update networks}
\State Sample mini-batch of $B$ instances from $\B_k$
\State Compute target actions and targets using \eqref{eqn:target_action} and \State \eqref{eqn:TD3_target} respectively
\State Update critic and actor based on \eqref{eqn:update-critic} and \eqref{eqn:TD3_actor_update}
\If{$t$ mod $policy\_delay$}
\State Update target networks using \eqref{eqn:Soft_update}
\EndIf
\EndIf
\EndFor
\end{algorithmic}
\end{algorithm}

\begin{figure}[t]
	\centering
	\includegraphics[width=0.48\textwidth]{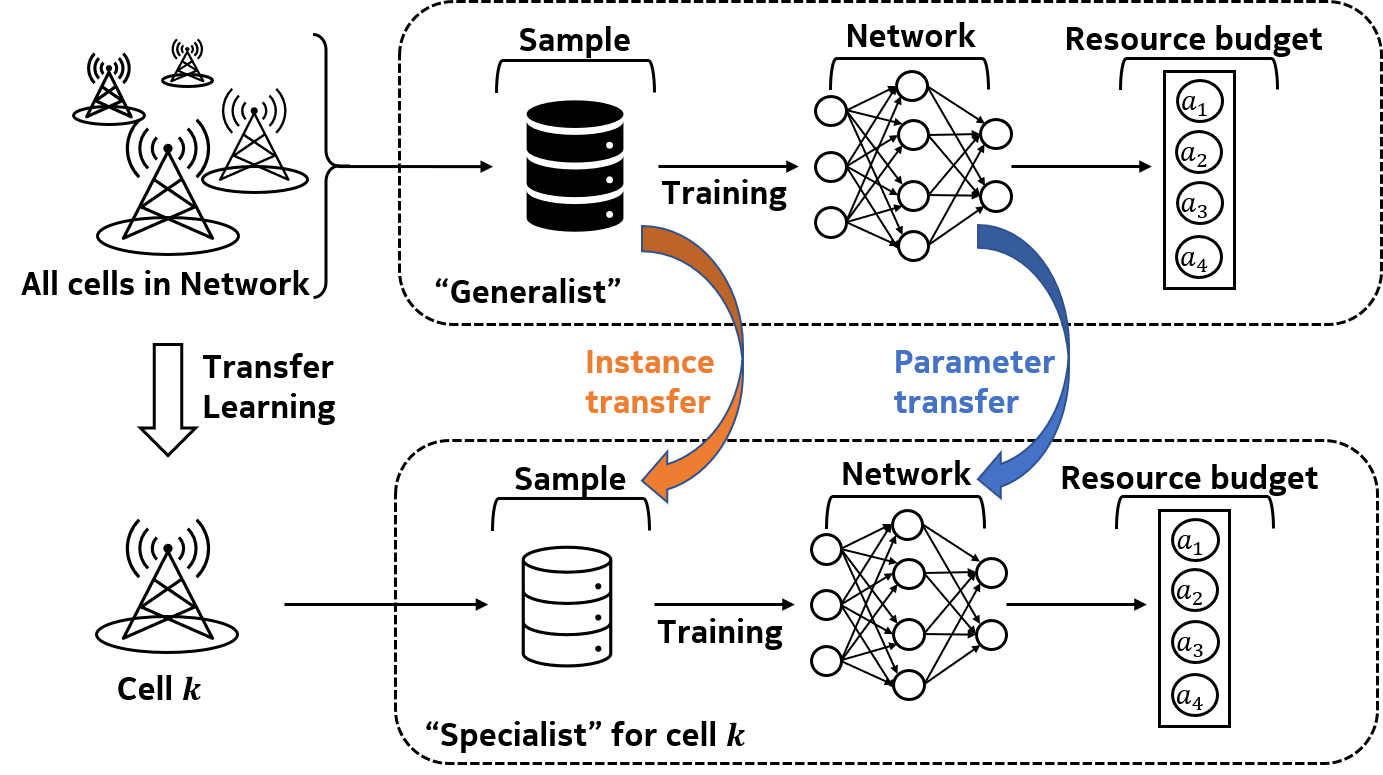}
	\caption{Generalist-to-Specialist transfer learning scheme}
	\label{fig:TLScheme}
\end{figure}

\section{Transfer Learning-Aided DIRP Algorithm}\label{sec:TL-MADRL}
As discussed in Section \ref{ssec:DIRP}, the DIRP algorithm achieves a good trade-off between reducing model complexity and capturing the inter-cell dependencies. 
However, each agent needs to learn the local policy from scratch and still faces the well-known challenge of the exploration-exploitation dilemma. The environment dynamics and state transitions are usually unknown at the early stage of training, and the agent cannot exploit its knowledge until the state-action space is exhaustively explored.
Moreover, because the local model is trained on a specific data domain, the learned model is sensitive to domain shift (a change in the data distribution between an algorithm's training dataset, and the dataset that it encounters when deployed). This means, even a slight change in the environment may result in deteriorated performance, and the agent may face a long period of retraining time.

To overcome the above-addressed challenges, we raise a hypothesis that some common hidden pattern may exist in the critic and actor networks across different agents, and propose to enhance the developed coordinated \ac{MADRL} algorithm with \ac{TL}. We expect the \ac{TL} to improve the model reproducibility and speed up the learning convergence by performing the following two major steps as demonstrated in Fig. \ref{fig:TLScheme}:
\begin{enumerate}
    \item \emph{Centralized training of a \lq\lq generalist\rq\rq}: A centralized controller collects the samples from all local agents $\Big(\big(\vs_k(t), \vc_k(t)\big), \va_k(t), \big(\vs_k(t+1), \vc_k(t+1)\big), r_k(t)\Big)$, $\forall k\in\K$ for a time period $t=0, \ldots, T^{\g}$ and trains a generalized model by interacting with the environment based on the same model in training for all agents.
    \item \emph{Distributed transfer learning and finetuning to the \lq\lq specialists\rq\rq:} After time slot $T^{\g}$, we transfer the learned knowledge in the \lq\lq generalist\rq\rq \ to each local agent (i.e., the \lq\lq specialists\rq\rq), and finetune the customized model locally. The details of different types of transferable knowledge are provided later in Section \ref{ssec:TLapproach}.
\end{enumerate}


\subsection{Transfer Learning Problem Formulation}
Before introducing the \ac{TL} problem in the context of \ac{MADRL}, let us first introduce a general definition of transfer learning. 

A \emph{domain} $\D:=\{\X, P(X)\}$ consists of a feature space $\X$ and its probability distribution $P(X), X\in\X$. A \emph{task} $\T:=\{\Y, f(\cdot)\}$ consists of a label space $\Y$ and a predictive function $f(\cdot)$, where $f(\cdot)$ can be written as $P(Y|X), Y\in\Y$ and $X\in\X$. Formally, the general definition of the \ac{TL} is given below.

\begin{definition}[Transfer Learning \cite{pan2009survey}] Given a source domain $\D_S$ and a source learning task $\T_S$, a target domain $\D_T$ and a target learning task $\T_T$, \ac{TL} aims to improve the learning of the target predictive function $f_T(\cdot)$ in $\D_T$ using the knowledge in $\D_S$ and $\T_S$, where $\D_S\neq\D_T$, or $\T_S\neq \T_T$.
\label{def:gen_TL}
\end{definition}

In the context of \ac{DRL}, a domain $\D:=\{\Ss, P(\vs)\}$ consists of the state space $\Ss$ and its probability distribution $P(\vs), \vs\in\Ss$, while the task $\T:=\{\A, \pi(\cdot)\}$ consists of the action space $\A$ and a policy function $\pi(\cdot)$. In general, the policy $\pi$ is a mapping from states to a probability distribution over actions. With the actor-critic method introduced in Section \ref{ssec:AC-method}, the policy directly maps the state space to optimized action, thus, we have $\pi:\Ss\to\A$.

In the scope of our proposed generalist-to-specialist \ac{TL}-DIRP algorithm, we introduce the following definitions of the source domain, source task, target domain, and target task. 
\begin{itemize}
\item {\bf Source domain}: $\D_S:=\D^{\g}$ consists of the joint state and communicated message space $\Ss^{\g}\times\R^N$ and its probability distribution $P\left(\vs^{\g}, \vc^{\g}\right)$, where $\vs^{\g}\in\Ss^{\g}:=\cup_{k\in\K} \Ss_k$ and $\vc^{\g}\in\R^N$. The state $\vs^{\g}$ and message $\vc^{\g}$ are collected by the centralized controller from all local agents. 
\item {\bf Source task}: $\T_S:=\T^{\g}$ consists of the general action space $\A^{\g}$ and the policy function $\pi^{\g}:\Ss^{\g}\times\R^N\to\A^{\g}$. The general policy $\pi^{\g}$ is trained on the instances collected by all agents. 
\item {\bf Target domain}: $\D_T:=\D_k^{\s}, k\in\K$ consists of the joint local state and communication message space $\Ss_k\times\R^N$ and its probability distribution $P\left(\vs_k, \vc_k\right)$, where $\vs_k\in \Ss_k$ and $\vc_k\in\R^N$.
\item {\bf Target task}: $\T_T := \T_k^{\s}, k\in\K$ consists of the local action space $\A_k$ and local policy $\pi_k:\Ss_k\times\R^N\to\A_k$.
\end{itemize}

The problem of \ac{TL} from a source \ac{DRL} agent as a \lq\lq generalist\rq\rq to a set of target \ac{DRL} agents, i.e., the local \lq\lq specialists\rq\rq, is formulated in Problem \ref{prob:TLinMADRL}. 

\begin{problem}\label{prob:TLinMADRL}
Given source domain $\D^{\g}:=\left\{\Ss^{\g}\times\R^N, P\left(\vs^{\g}, \vc^{\g}\right)\right\}$ and pretrained source task $\T^{\g}:=\left\{\A^{\g}, \pi^{\g}(\cdot)\right\}$, transfer learning aims to learn an optimal local policy for the target domain $\D_k^{\s}:=\left\{\Ss_k\times\R^N, P\left(\vs_k, \vc_k\right)\right\}$, $\forall k\in\K$ by leveraging the knowledge extracted from $\left(\D^{\g}, \T^{\g}\right)$, as well as the knowledge exploited in the target domain $\D_k^{\s}$. The problem is given by
\begin{align}
\max\limits_{\pi_k|\pi_k^{(0)}= \Lambda\left(\pi^{\g}\right)} \ & \Ex_{\pi_k} \left[\sum_{t=0}^{T} \gamma_k^t \tilde{r}_k\big(\vs_k(t), \vc_k(t), \va_k(t) \big)\right] \label{eqn:TL_prob}\\
\mbox{s.t. } & (\vs_k, \vc_k, \va_k) \in \Omega\left(\D^{\g}, \D_k^{\s}, \A^{\g}, \A_k\right).\nonumber
\end{align}
where $\Lambda\left(\pi^{\g}\right)$ is the \emph{policy transfer strategy} which maps the pretrained source policy $\pi^{\g}$ to an initial local policy $\pi_k^{(0)}$, while $\Omega\left(\D^{\g}, \D_k^{\s}, \A^{\g}, \A_k\right)$ is the \emph{instance transfer strategy} which extracts the instances from the source domain and combines them with the experienced instances from the target domain. 
\end{problem}

\subsection{Transfer Learning Approaches}\label{ssec:TLapproach}
The problem defined in Eq.~\eqref{eqn:TL_prob} offers various options for transferable knowledge:
\begin{itemize}
    \item \emph{Pretrained model transfer}: The policy transfer strategy $\Lambda(\cdot)$ simply maps the pretrained source policy to itself, i.e., the local agent uses the pretrained general policy $\pi^{\g}$ as the initial policy $\pi_k^{(0)}$ and finetunes it by further interacting with the environment with locally made decisions.
    \item \emph{Feature extraction}: $\Lambda(\cdot)$ keeps partial knowledge of $\pi^{\g}$. In \ac{DRL}, the policy $\pi^{\g}\left(\vs^{\g}, \vc^{\g}|\ve{\phi^{\g}}\right)$ is characterized by the pretrained parameters (weights) of the neural networks. Feature extraction freezes partial of the layers (usually the lower layers) of the pretrained neural networks while leaving the rest of them to be randomly initialized.
    \item \emph{Instance transfer}: Except for the instances from the target domain, the agent also trains its policy using the extracted instances from the source domain. The instance transfer strategy $\Omega\left(\cdot\right)$ decides which instances are chosen from the source domain to be combined with the instances from the target domain in the local replay buffer.
\end{itemize}

The above-mentioned knowledge from the source domain and task can be transferred separately or in a combined manner. In this paper, we focus on studying the following three \ac{TL} schemes:
\begin{itemize}
\item {\bf Pretrained model transfer only}: Each local agent $k$ uses the pretrained general policy $\pi^{\g}$ to initialize the local policy $\pi_k^{(0)}$. With the actor-critic method described in Section \ref{ssec:AC-method}, we simply load the pretrained parameters of the actor and critic networks from the generalist to the local agents. However, when the difference between the source and target domain is large, the local agent still needs extensive exploration to finetune the general policy to a customized local policy. 
\item {\bf Instance transfer only}: Each local agent offloads a set of selected instances in the source domain from the centralized controller to the local replay buffer. Then, the local agent trains a policy from scratch with the replay buffer containing mixed offline instances from the source domain and the experienced online instances in the target domain. In this paper, we select the instances collected from the exact same local agent. Future work includes the similarity analysis between agents and instance selection from similar agents, which falls into the subject of \emph{domain adaptation} \cite{kang2019contrastive}. 
\item {\bf Combined model and instance transfer}: To fully exploit the transferable knowledge, we combine the pretrained model transfer and instance transfer. Firstly, each local agent retrieves $\pi^{\g}$ from the centralized controller and uses it to initialize the local policy $\pi_k^{(0)}$. Then, we further investigate two options for local finetuning:
\begin{itemize}
    \item \emph{Online finetuning with mixed replay buffer}: The local agent further online finetunes the policy with the replay buffer containing both the offloaded instances from the source domain and the locally experienced instances from the target domain. 
    \item \emph{Offline finetuning with offloaded instances \& online finetuning with experienced instances}: The local agent first offline finetunes $\pi^{\g}$ with the offloaded instances. Then, the offline finetuned model is used to initialize $\pi_k^{(0)}$ and further finetuned online with the locally experienced instances in the target domain. 
\end{itemize}
\end{itemize}

Note that our experiments focus on the pretrained model transfer and instance transfer, while do not include the feature extraction. This is because, feature exaction usually performs well when the target domain is highly similar to the source domain. However, in general, the similarity between the generalist's domain and the specialist's domain is not sufficiently high. Thus, the feature exaction method may better suit the scenario of inter-agent \ac{TL}, while it may not be appropriate for generalist-to-specialist knowledge transfer.

We illustrate the \ac{TL}-DIRP algorithm with a combined model and instance transfer in Algorithm \ref{algo:TL-DIRP}.

\begin{algorithm}
\caption{Transfer learning-Aided DIRP Algorithm}
\label{algo:TL-DIRP}
\begin{algorithmic}[1]
\State {\bf \RNum{1}. Generalist training
in centralized controller}
\State Initialize generalist's critics $Q_{\theta_1^{\g}}$, $Q_{\theta_2^{\g}}$ and actor $\pi_{\phi^{\g}}$ with random parameters $\theta_1^{\g}$, $\theta_2^{\g}$, $\phi^{\g}$
\State Initialize target networks $\theta_1'^{\g}\leftarrow\theta_1^{\g}$, $\theta_2'^{\g}\leftarrow\theta_2^{\g}$, $\phi'^{\g}\leftarrow\phi^{\g}$
\State Initialize empty replay buffer $\B^{\g}$
\State Define time periods $\Ho^{\g}, \Ho^{\s}$ for generalist training and specialist finetuning respectively
\For{$t\in\Ho^{\g}$}
\State Collect observations of local states $\vs_k(t)$ and \State received information $\vc_k(t)$, $\forall k\in\K$
\State Use general policy $\pi^{\g}$ to select and execute action \State $\va_k(t), \forall k\in\K$ 
\State Observe the next local states $\vs_k(t+1)$ and information \State $\vc_k(t+1)$, compute local rewards $r_k(t)$, $\forall k\in\K$
\State Store $K$ instances in replay buffer $\B^{\g}$
\State Train and update the general critics $Q_{\theta_i^{\g}}, i=1,2$, \State actor $\pi^{\g}$, and target critics $Q_{\theta_i^{'\g}}, i=1,2$ and actor \State $\pi_{\phi^{'\g}}$ using the TD3 algorithm in Section \ref{ssec:AC-method} 
\EndFor
\State {\bf \RNum{2}. Specialist finetuning in local agents}
\State Initialize parameters for critics $Q_{\theta_1^k}$, $Q_{\theta_2^k}$ and actor $\pi_{\phi^k}$ with $\theta_1^k\leftarrow\theta_1^{\g}$, $\theta_2^k\leftarrow \theta_2^{\g}$, $\phi^k\leftarrow \phi^{\g}$, $\forall k\in\K$
\State Initialize target networks $\theta_1'^{k}\leftarrow\theta_1^k$, $\theta_2'^k\leftarrow\theta_2^k$, $\phi'^k\leftarrow\phi^k$
\State Offload selected instances from $\B^{\g}$ to $\B_k$ 
\For{$t\in\Ho^{\s}$}
\For{Local agent $k\in K$}
\State Finetune local policy with Algorithm \ref{algo:TD3-DIRP} (except for \State the initialization steps)
\EndFor
\EndFor
\end{algorithmic}
\end{algorithm}


\section{Performance Evaluation}\label{sec:simu}
In this section, we evaluate the performance of the proposed methods for inter-cell slicing resource partitioning introduced in Sections \ref{sec:MADRL} and \ref{sec:TL-MADRL} with a system-level simulator \cite{SeasonII}, which mimics real-life network scenarios with customized network slicing traffic, user mobility, and network topology.

\begin{figure}[t]
    \centering
    \includegraphics[width=.45\textwidth,height=4.5cm]{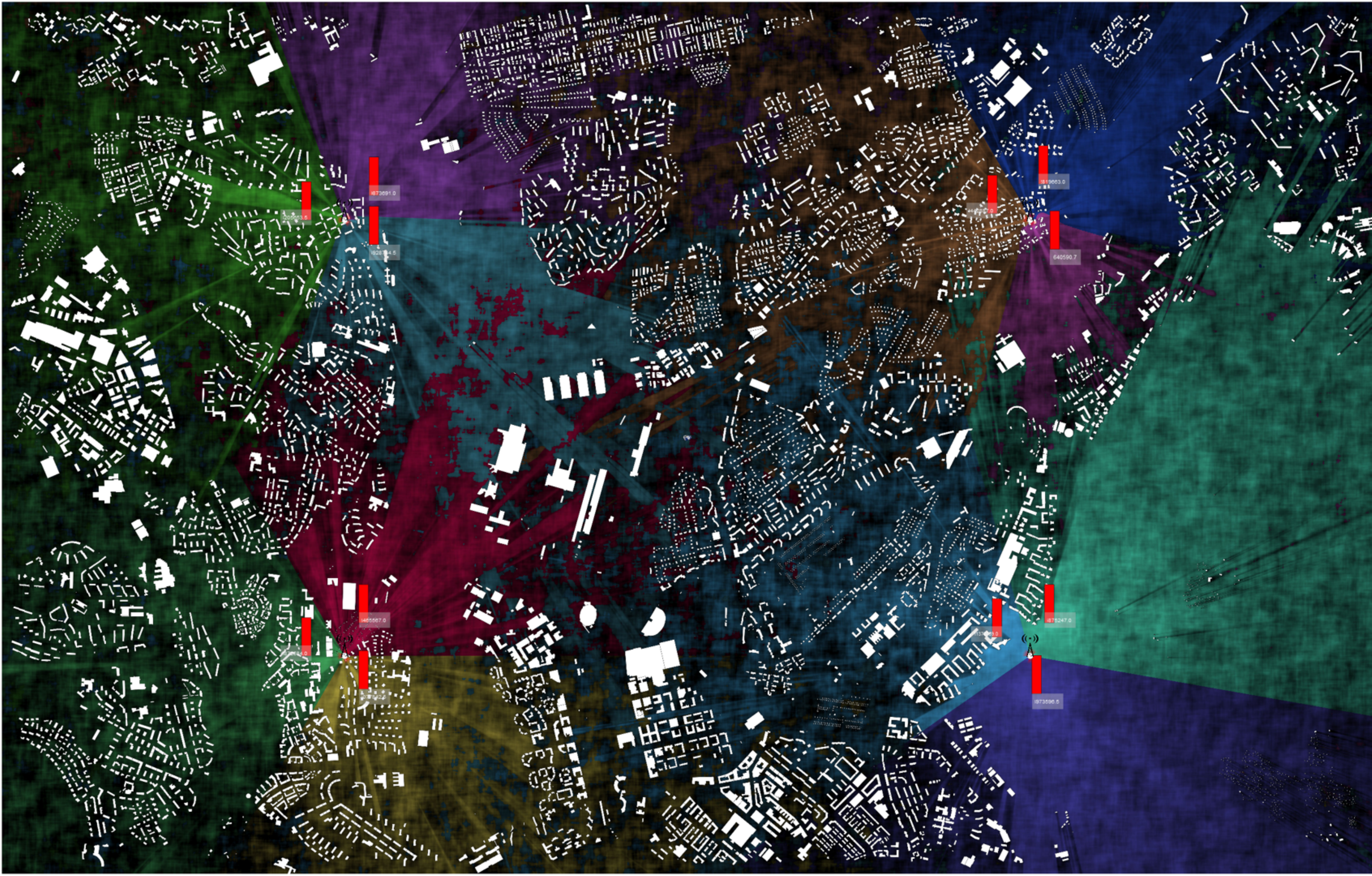}
    \caption{Network environment setup with $12$ cells}
    \label{fig:SeaEnv}
\end{figure}

To implement our proposed \ac{DRL} solution, we build in the simulator a network with $4$ sites ($12$ cells) covering an urban area of Helsinki city, as demonstrated in Fig. \ref{fig:SeaEnv}, consisting of $4$ three-sector macro sites. All cells are deployed using LTE radio technology with $2.6$ GHz. We use the realistic radio propagation model Winner+ \cite{Winner}.

The network is built up with $N=4$ network slices, with per-slice throughput requirements of $\phi_1^* = 4$ MBit/s, $\phi_2^* = 1$ MBit/s, $\phi_3^* = 3$ MBit/s, and $\phi_4^* = 0.5$ MBit/s and per-slice delay requirements of $d_1^* = 1$ ms, $d_2^* = 1.5$ ms, $d_3^* = 2$ ms, and $d_4^* = 1$ ms respectively. All cells in the network have a fixed bandwidth of $20$ MHz.

\begin{figure}[t]
    \centering
    \includegraphics[width=.45\textwidth]{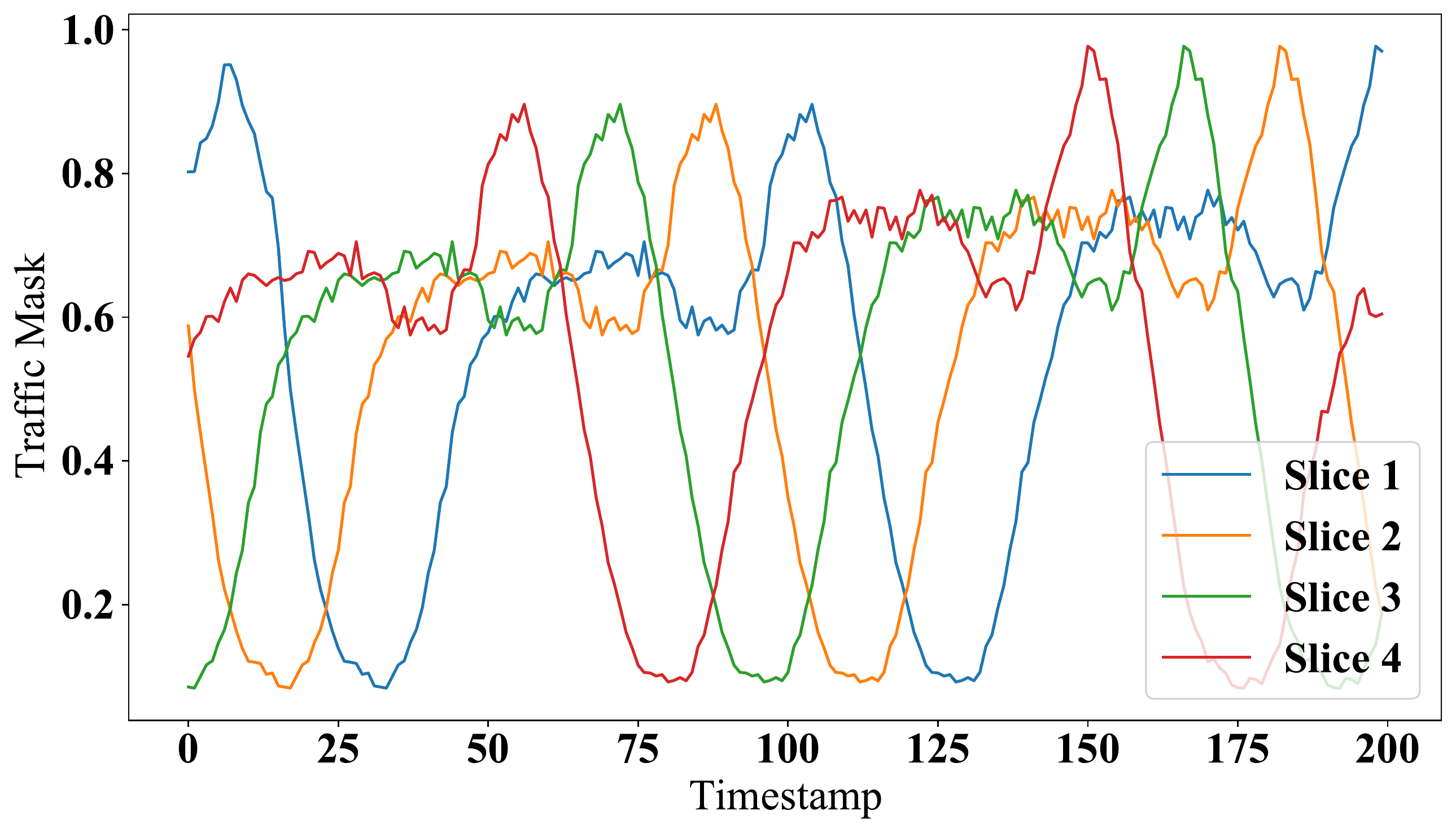}
    \caption{The first two days of a three-week traffic mask}
    \label{fig:TrafMask}
\end{figure}

We define four groups of \acp{UE} associated with each defined slice respectively, i.e., $16$ groups of \acp{UE} in total, all with the maximum group size of $10$. \acp{UE} are moving uniformly randomly within the defined moving sphere of each group. The positions and moving radius of \acp{UE} groups are defined heterogeneously to ensure that each site can serve \ac{UE} from all slices. To imitate the time-varying traffic pattern, we also apply a time-dependent traffic mask $\tau_n(t)\in[0, 1]$ for each slice $n\in\N$ to scale the total number of \acp{UE} in the scenario. In Fig. \ref{fig:TrafMask}, we demonstrate the changes of the first 2 days of a three-week traffic mask. The \ac{UE} traffic volume is updated every timestamp, which corresponds to $15$ minutes in real time, also known as the typical \ac{KPI} reporting time in \ac{OAM}. In the experiments, the entire traffic mask is extended and periodically repeated after every $2016$ timestamps corresponding to the three-week time period ($96$ timestamps per day).

\subsection{Schemes and Baselines to Compare}
\label{ssec:comparedSchemes}
For performance evaluation, we compare the proposed \ac{DIRP} and \ac{TL}-\ac{DIRP} algorithms with the following three baselines: 

\begin{table*}[ht]
\caption{Comparison of Dimensions of \ac{DRL} Models Used in Simulation}
\label{Table_RL_Compare}
\begin{tabular*}{1.02\textwidth}{@{\extracolsep{\stretch{1}}}*{5}{r}@{}}
\cline{1-4}
\multicolumn{1}{|l|}{}      & \multicolumn{1}{l|}{{\bfseries BL-Cen}}         & \multicolumn{1}{l|}{{\bfseries BL-Dist}}    & \multicolumn{1}{l|}{{\bfseries DIRP \& TL-DIRP}}     &  \\ \cline{1-4}
\multicolumn{1}{|l|}{{\bfseries State}}  & \multicolumn{1}{l|}{Global state $\vs\in\R^{240}$} & \multicolumn{1}{l|}{Local state $\vs_k\in\R^{20}$ }    & \multicolumn{1}{l|}{Local state with extracted message $[\vs_k, \vc_k]\in\R^{24}$} &  \\ \cline{1-4}
\multicolumn{1}{|l|}{{\bfseries Action}} & \multicolumn{1}{l|}{Global action $\va\in [0,1]^{48}$}       & \multicolumn{1}{l|}{Local action $\va_k\in[0,1]^{4}$}   & \multicolumn{1}{l|}{Local action $\va_k\in[0,1]^{4}$}                       &  \\ \cline{1-4}
\multicolumn{1}{|l|}{{\bfseries Reward}} & \multicolumn{1}{l|}{Global reward $r_G^m$ in Eq.~\eqref{eqn:global_rewardfunction_maxmin}}       & \multicolumn{1}{l|}{Local reward $r_k^m$ in Eq.~\eqref{eqn:local_rewardfunction_maxmin}} & \multicolumn{1}{l|}{Local reward $r_k^m$ in Eq.~\eqref{eqn:local_rewardfunction_maxmin}}                     &  \\ \cline{1-4}
                             &                                          &                                     &                                                         & 
\end{tabular*}
\end{table*}


\begin{itemize}
	\item {\bf BL-Cen}: centralized \ac{DRL} approach solving Eq.~\eqref{eqn:problem} referring to \cite{Li2018DeepRL}. We assume that a single agent has full observation of the global state $\vs\in\Ss$, computes the global reward and makes the decision of the slicing resource partitioning for all agents $\va\in\A$.
	\item {\bf BL-Dist}: distributed \ac{DRL} approach without inter-agent coordination referring to \cite{Zhao2019DeepRL}.
	\item {\bf BL-Heur}: a traffic-aware heuristic approach that assumes perfect knowledge about per-slice traffic demand, and dynamically adapts to the current per-slice traffic amount. It is implemented by dividing the resource in each cell $k\in\K$ to each slice proportionally to the amount of traffic demand per slice. 
\end{itemize}

The \ac{DRL}-based schemes to evaluate and compare are summarized in Table \ref{Table_RL_Compare}. 

Similarly, to evaluate the \ac{TL}-DIRP algorithm and compare between different types of knowledge to transfer, we implement the proposed \ac{TL} method in Section \ref{sec:TL-MADRL}, i.e., centralized training of a generalist and then distributed finetuning to specialist. We compare different transferable knowledge: instances, pretrained model, and combined instances and pretrained model. In addition, to ensure a safer exploration and better performance during online training, we perform the offline finetuning using the transferred instance before the online training in each local agent. 

\begin{itemize}
	\item {\bf Gen}: centralized training of a general policy in the centralized controller based on the collected samples from all local agents, as described in subsection \ref{algo:TL-DIRP}.
	\item {\bf Spec}: distributed finetuning of the specialists with full knowledge transfer. Each local agent initializes its critic and actor networks with the generalist's model parameters. It also initializes the local replay buffer with the offloaded selected instances from the generalist's buffer. 
	\item {\bf Spec-Instance}: distributed finetuning of the specialists with instance transfer only. The model parameters in each local agent are randomly initialized.
	\item {\bf Spec-Model}: distributed finetuning of the specialists with model transfer only. Each local agent initializes its critic and actor networks by loading the generalist's model parameters, while the local buffer is initialized as an empty queue.
	\item {\bf TL-DIRP}: In addition to Spec (full knowledge transfer), we apply the offline finetuning based on the transferred instances before the online training.
\end{itemize}

Note that for \lq\lq generalist-to-specialist" \ac{TL} schemes with complete knowledge we apply both max-min fairness and logarithmic utilities as local reward $r_k$ for $k\in\K$ respectively, as:

\begin{itemize}
    \item {\bf TL-DIRP-Maxmin}: TL-DIRP approach with max-min fairness reward bases on Eq.~\eqref{eqn:local_rewardfunction_maxmin}.
    \item {\bf TL-DIRP-Log}: TL-DIRP approach with on logarithmic utility reward based on Eq.~\eqref{eqn:local_rewardfunction_log}.
\end{itemize}


\subsection{Hyperparameters used for Learning}
As for \ac{DRL} training, we use \ac{MLP} architecture for actor-critic networks of \ac{TD3} algorithm. In BL-Cen scheme, the models of the actor and critic networks are both built up with $3$ hidden layers, with the number of neurons $(384, 192, 64)$ and $(324, 144, 64)$, respectively. While for BL-Dist and DIRP schemes, both actor-critic networks only have $2$ hidden layers, with the number of neurons $(48, 24)$ and $(64, 24)$, respectively. In all schemes, the learning rate of actor and critic are $0.0005$ and $0.001$ respectively with Adam optimizer and training batch size of $32$. 
We choose a small \ac{DRL} discount factor $\gamma = 0.1$, since the current action has a strong impact on the instantaneous reward while a weaker impact on the future reward. For the distributed \ac{DRL} approaches, we only apply $100$ steps for exploration, while for the centralized approaches we apply $500$ steps of exploration, since the centralized agent has much higher dimensions of state and action. After the exploration phase, we apply $5000$ steps for training, and the final $500$ steps for evaluation of all approaches. 

In \ac{TL} training, we apply the same \ac{DRL} settings. For \ac{TL} training setup, we set $100$ steps for exploration, $5000$ steps for learning, and $500$ steps for evaluation in Gen and Spec-Model schemes, while in other \ac{TL} execution schemes, we skip the exploration phase. The result of each process is derived from the average of $3$ times of experiments.

In this work, we apply an orientated exploration strategy that chooses the new action under the recommendation of the traffic-aware heuristic policy, namely, the heuristic baseline BL-Heur. The reason is that we observe that BL-Heur provides sub-optimal performance without any training process. At the beginning of the exploration phase,  the probability of using traffic-aware exploration is $0.5$, and that of random exploration is also $0.5$. Then, during the exploration, the probability of traffic-aware exploration gradually increases, and that of random exploration decreases.


\subsection{Performance Comparison}

\begin{figure}[t]
	\centering
	\includegraphics[width=.45\textwidth]{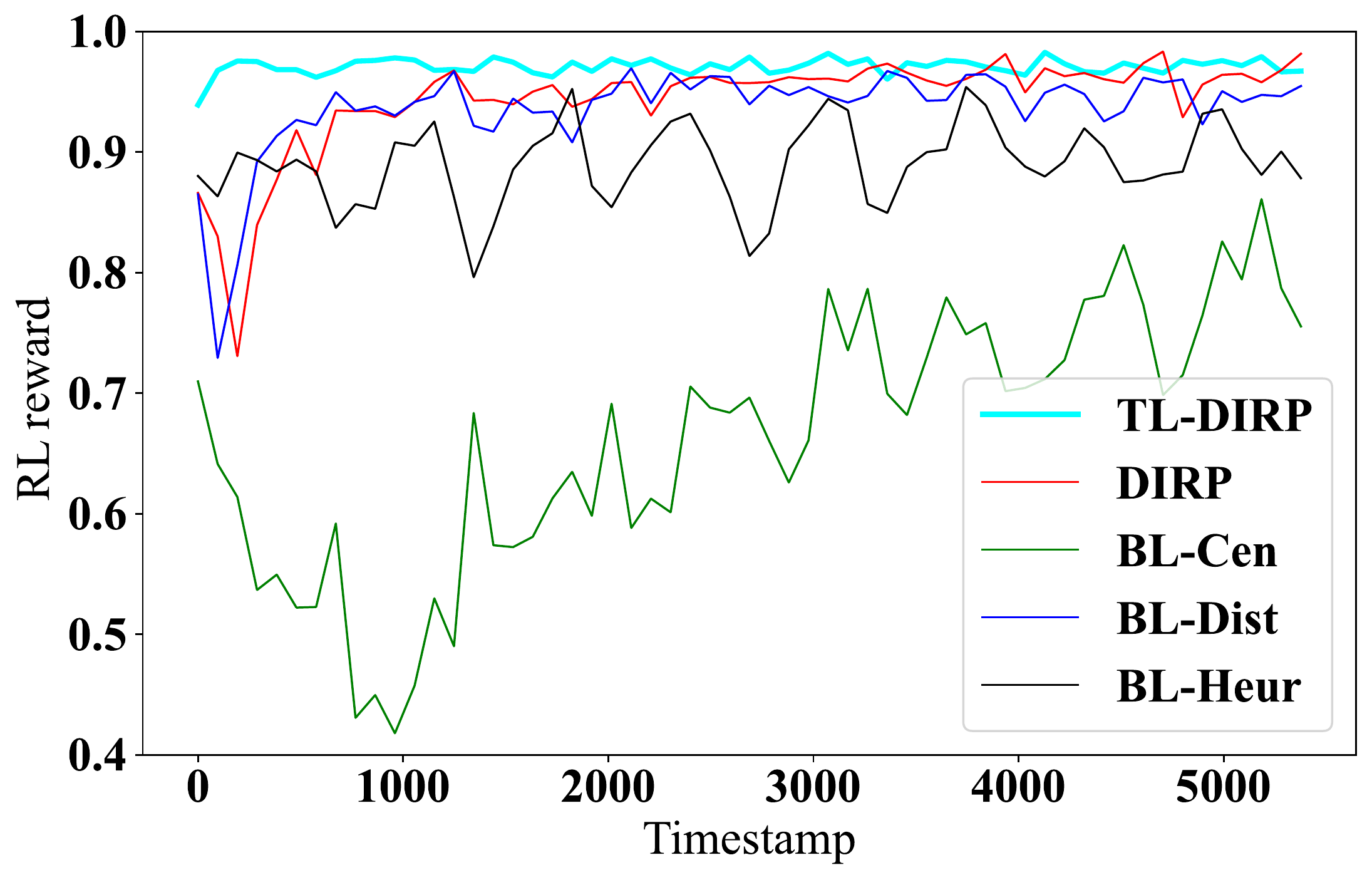}
	\caption{Comparison of reward among schemes}
	\label{fig:Compare_baseline_reward}
\end{figure}

\subsubsection{Comparison of the Distributed \ac{MADRL} Schemes}
In this comparison, we apply the reward design for max-min fairness to all approaches, i.e., global reward based on Eq.~\eqref{eqn:global_rewardfunction_maxmin} for BL-Cen and local reward based on Eq.~\eqref{eqn:global_rewardfunction_log} for BL-Dist and DIRP. While for comparison between the different reward functions, we implement DIRP algorithm with both types of local reward $r_k$ based on Eq.~\eqref{eqn:local_rewardfunction_maxmin} and Eq.~\eqref{eqn:local_rewardfunction_log}.

Fig. \ref{fig:Compare_baseline_reward} demonstrates the comparison of max-min fairness reward Eq.~\eqref{eqn:global_rewardfunction_maxmin} during the training process among the baseline schemes BL-Cen, BL-Dist, BL-Heur, and the proposed DIRP and TL-DIRP algorithms.

As shown in Fig. \ref{fig:Compare_baseline_reward}, TL-DIRP provides the best performance among all approaches in terms of faster convergence, higher start point, and higher robustness after convergence.



While in comparison to baselines, \ac{DIRP} algorithm achieves significantly better global reward than BL-Heur after convergence. Note that BL-Heur is already a well-performed baseline because it assumes perfect traffic awareness and offers all resources to the \acp{UE}. On the other hand, BL-Cen fails to achieve performance as good as \ac{DIRP} within the same training time. As Table \ref{Table_RL_Compare} indicates, the dimensions of the state and action spaces of BL-Cen are much higher than the distributed approaches, making the training process more difficult for large-scale networks. Not only converges BL-Cen slower, but it also often experiences poor performance at the early stage of training. The training curves are turbulent, corresponding to the time-varying traffic demand in Fig. \ref{fig:TrafMask}, while \ac{DIRP} is more robust compared to BL-Heur and BL-Cen.

\begin{figure}[t]
	\centering
	\includegraphics[width=.45\textwidth]{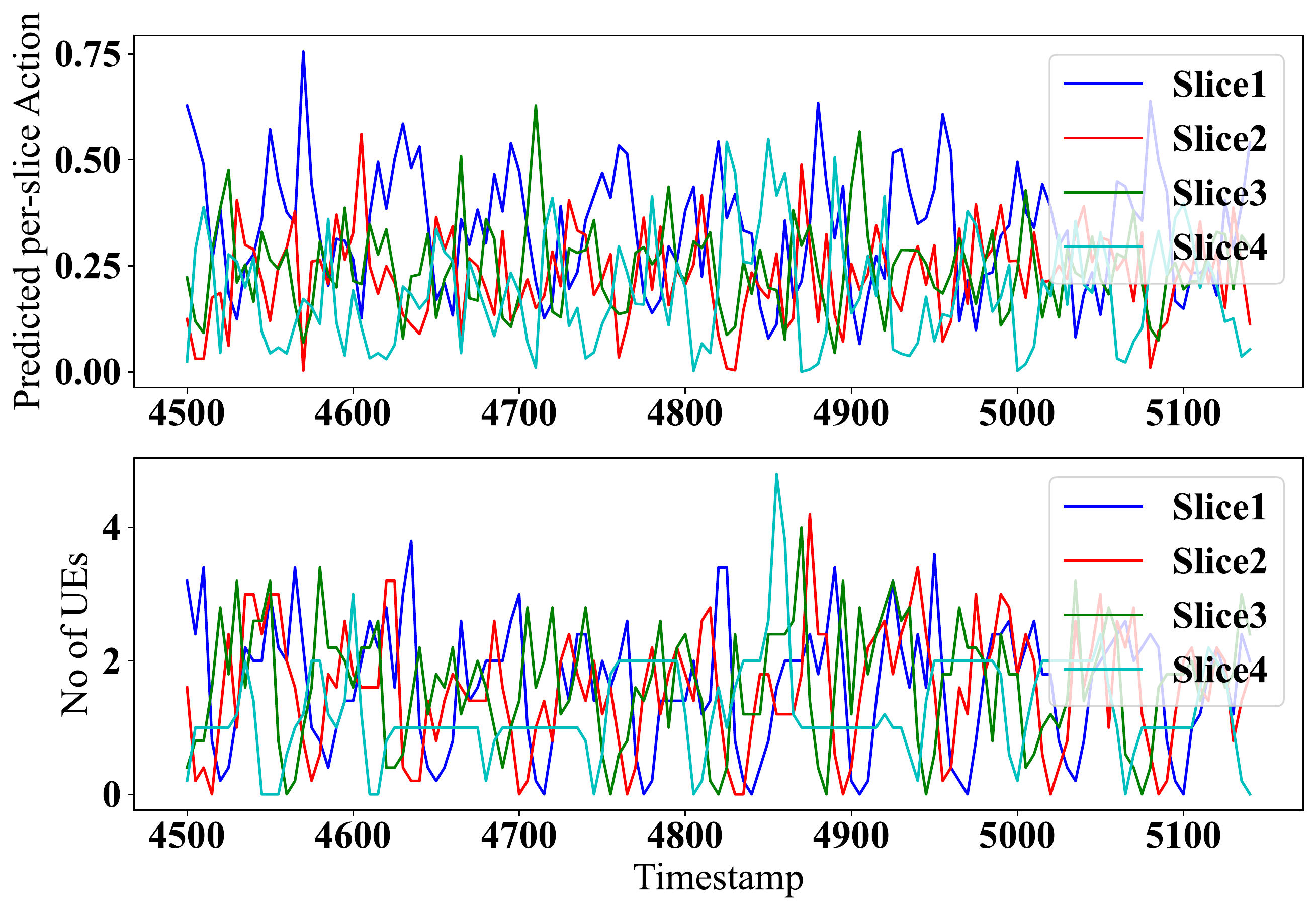}
	\caption{Adaptive action to traffic mask after training}
	\label{fig:Action_UE}
\end{figure}

In comparison between the two distributed schemes, according to Fig. \ref{fig:Compare_baseline_reward}, DIRP outperforms BL-Dist scheme within the same training time period in terms of both converged global reward and convergence rate, which verifies the advantage of inter-agent coordination.


Fig. \ref{fig:Action_UE} shows the predicted action, i.e., per-slice resource partitioning as the ratio, and the actual traffic amount of \ac{DIRP} in cell $k=5$ after convergence. it verifies that the \ac{DRL} approach well adapts its predicted actions to the dynamic network traffic demand with respect to different slice-specific \ac{QoS} requirements.

\begin{figure}[t]
	\centering
	\includegraphics[width=.45\textwidth]{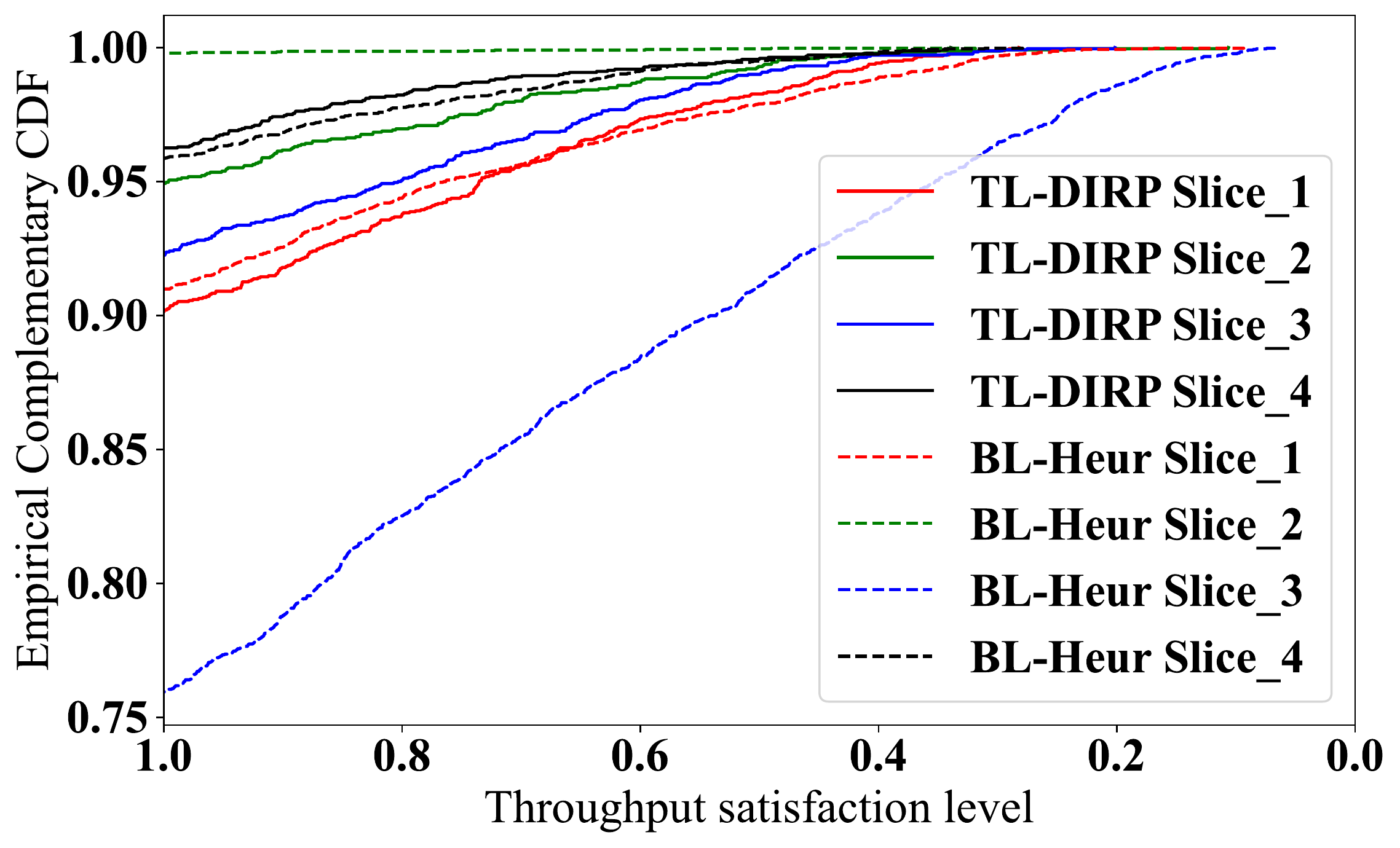}
	\caption{Comparing throughput \ac{QoS} between TL-DIRP and BL-Heur}
	\label{fig:DRL_thr_CDF}
\end{figure}

\begin{figure}[t]
	\centering
	\includegraphics[width=.45\textwidth]{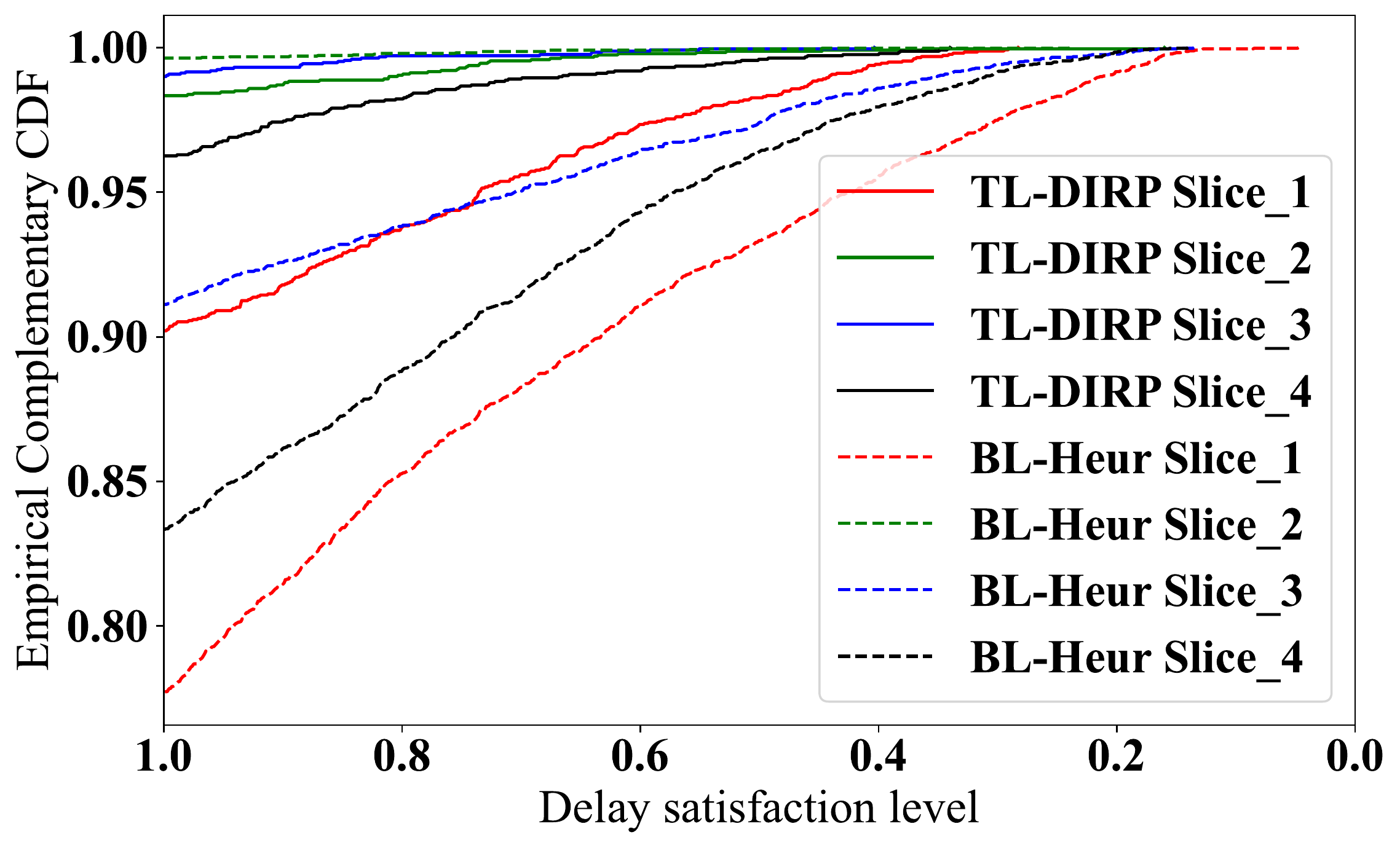}
	\caption{Comparing delay \ac{QoS} between TL-DIRP and BL-Heur}
	\label{fig:DRL_delay_CDF}
\end{figure}
Although \ac{DIRP} shows better performance than baselines, it still faces two major challenges: slow convergence and oscillation. In Fig. \ref{fig:Compare_baseline_reward}, we show that \ac{TL}-\ac{DIRP} overcomes these challenges by transferring prelearned knowledge.  In particular, TL-DIRP achieves a much higher reward from the beginning of the learning process and quickly converges after a few hundred timestamps, while \ac{DIRP} converges much slower because each local agent needs to learn from scratch. \ac{TL}-\ac{DIRP} outperforms \ac{DIRP} in terms of both convergence rate and converged performance within the same time period.

Fig. \ref{fig:Compare_baseline_reward} shows the evolving algorithms' performance during the training and testing process, while in the following, let us take a deeper look into the distributions of the converged service quality in terms of throughput and delay satisfaction level for each slice. 
Fig. \ref{fig:DRL_thr_CDF} and Fig. \ref{fig:DRL_delay_CDF} illustrate the empirical complementary \ac{CDF} (or called survival function) which equals $1-F_X(x)$, where $F_X(x)$ denotes the \ac{CDF} of per-slice throughput and delay satisfaction level between \ac{TL}-\ac{DIRP} and BL-Heur schemes, respectively.

Fig. \ref{fig:DRL_thr_CDF} shows that \ac{TL}-\ac{DIRP} achieves $14\%$ higher the worst-case throughput \ac{QoS} among all slices than the traffic-aware baseline BL-Heur. It also guarantees that all the slices achieve a throughput satisfaction level above $90\%$, while BL-Heur serves Slice $3$ with only $~75\%$ throughput satisfaction level.

Similar observation can be made for the delay satisfaction level in Fig. \ref{fig:DRL_delay_CDF}. \ac{TL}-\ac{DIRP} provides over $90\%$ of the delay satisfaction level for all slices, while BL-Heur serves Slice $1$ and $3$ with only $~77\%$ and $~83\%$ respectively. In terms of the average delay satisfaction level over all slices, TL-DIRP achieves over $96\%$ while BL-Heur only $88\%$. We observe that \ac{TL}-\ac{DIRP} attempts to fulfill more critical requirements by compromising resources from the less demanding slices while remaining sufficient satisfaction levels in others.


\begin{figure}[t]
	\centering
	\includegraphics[width=.45\textwidth]{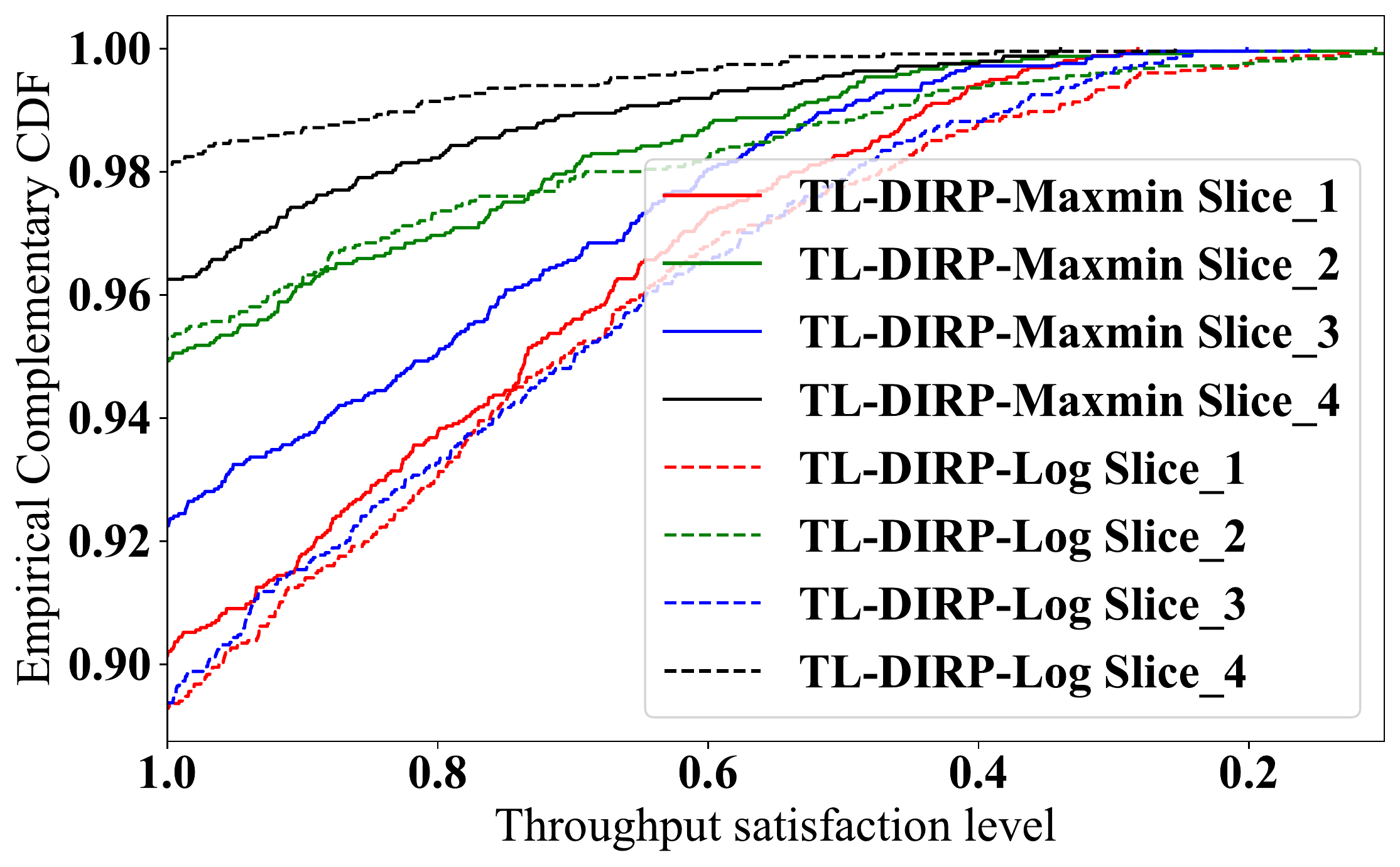}
	\caption{Comparing throughput \ac{QoS} between utilities}
	\label{fig:Utility_thr_CDF}
\end{figure}

\begin{figure}[t]
	\centering
	\includegraphics[width=.45\textwidth]{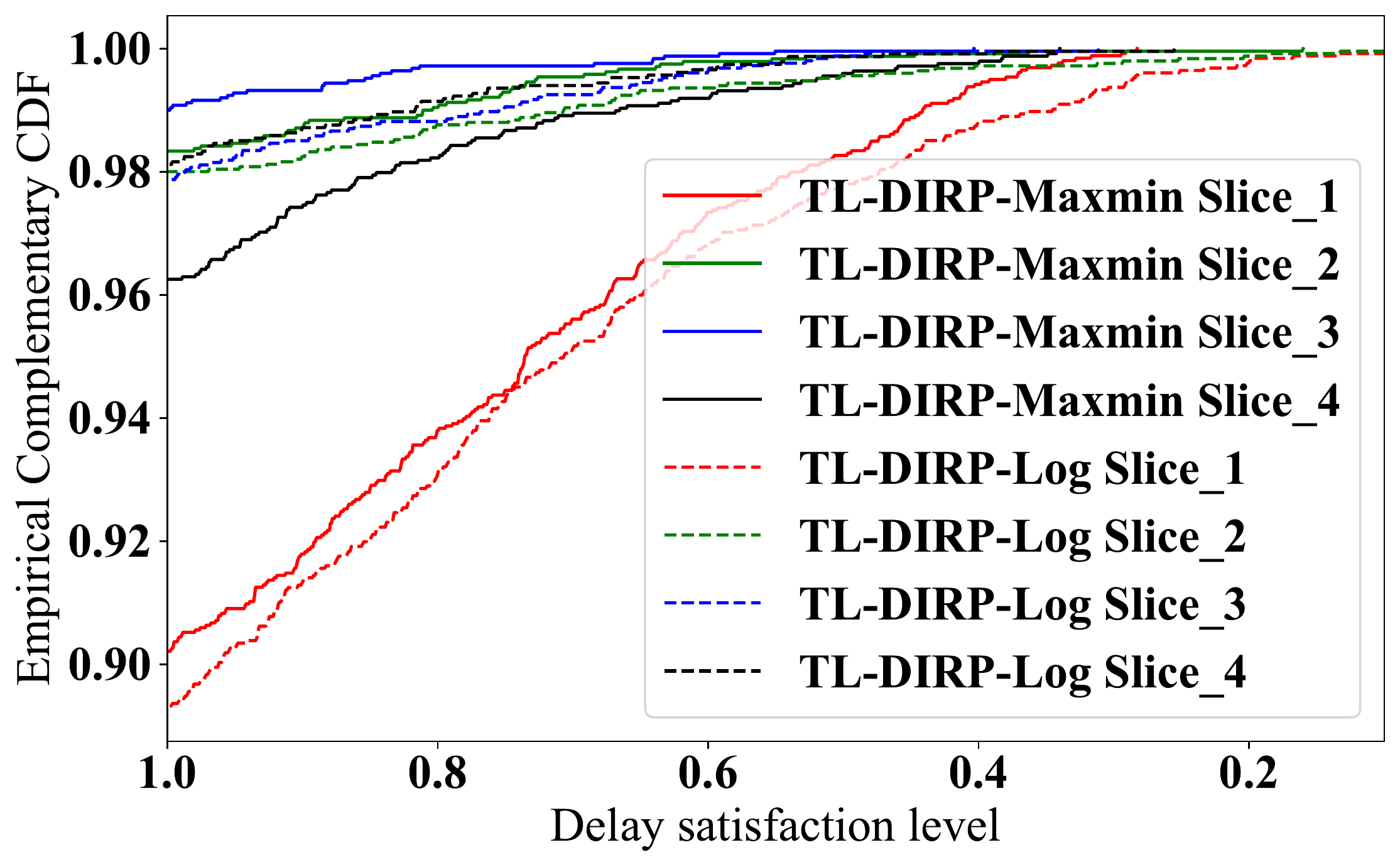}
	\caption{Comparing delay \ac{QoS} between utilities}
	\label{fig:Utility_delay_CDF}
\end{figure}

\subsubsection{Comparison between reward function with two utilities}

Fig. \ref{fig:Utility_thr_CDF} and Fig. \ref{fig:Utility_delay_CDF} compare the two designs of the reward function, corresponding to max-min fairness Eq.~\eqref{eqn:local_rewardfunction_maxmin} and maximizing average logarithmic utilities Eq.~\eqref{eqn:local_rewardfunction_log}, respectively.
They demonstrate the empirical complementary \ac{CDF} of \ac{QoS} in terms of throughput and delay satisfaction level for \ac{TL}-\ac{DIRP} with both reward functions. The results show that max-min fairness gives the maximum protection to the slice with the weakest performance, such that the minimum per-slice satisfaction level over all slices achieves $90\%$ for both throughput and delay, while maximizing average logarithmic utilities provides slightly lower satisfaction levels, about $89\%$ for both throughput and delay, but higher maximum per-slice throughput satisfaction levels. This is because the logarithmic utility tends to distribute the resource more efficiently than max-min fairness, i.e., allocating more resources to the slice that can improve the averaged performance over all slices.

From an engineering perspective, max-min fairness is preferred for scenarios that require sufficiently good performance for all slices, especially those highly demanding ones. While the logarithmic utility is more suitable for cases that desire higher resource efficiency.

\subsubsection{Comparison of the Transfer Learning Methods}

\begin{figure}[t]
	\centering
	\includegraphics[width=.45\textwidth]{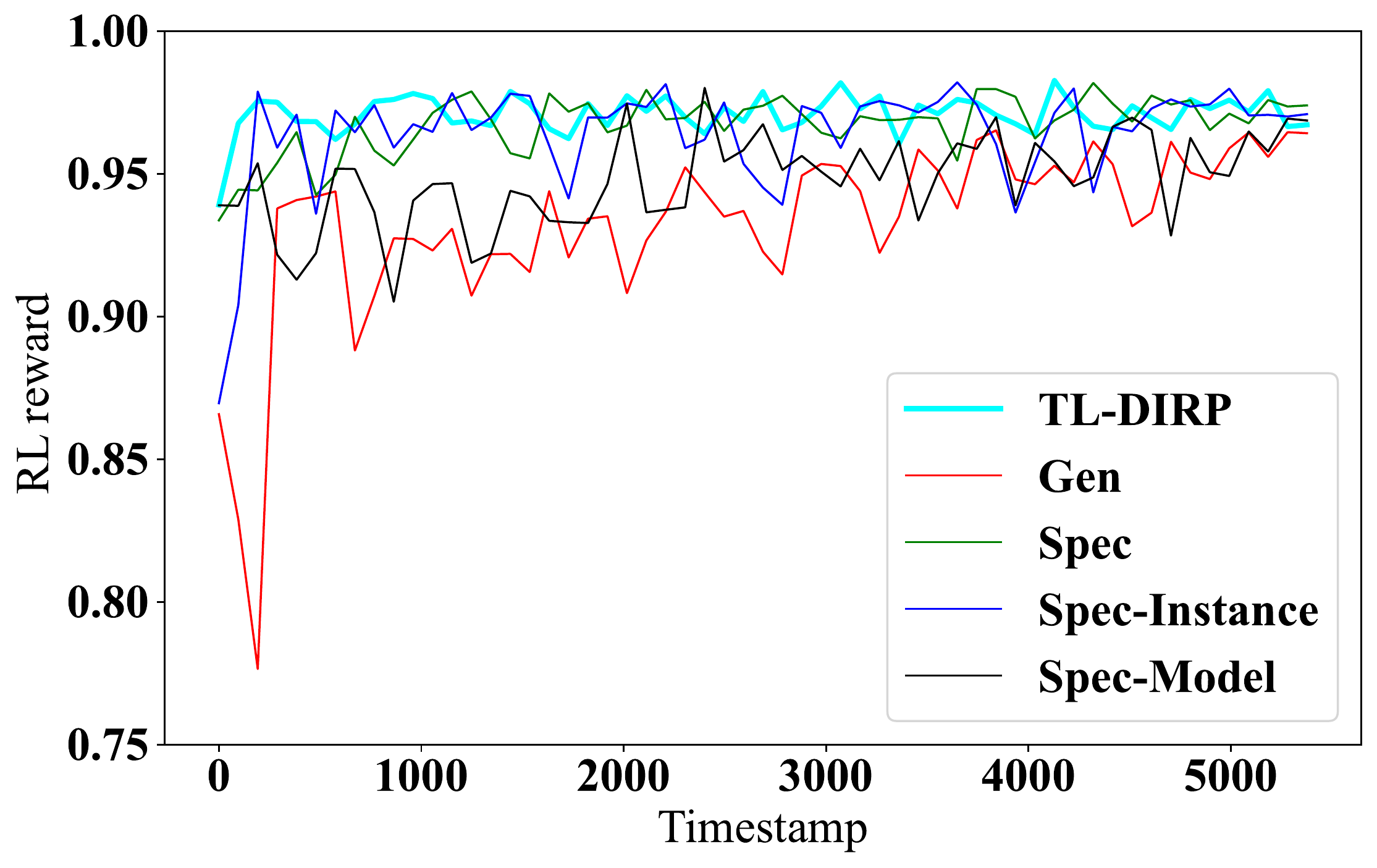}
	\caption{Comparing of reward among TL schemes}
	\label{fig:TL_Reward_Compare}
\end{figure}

\begin{table}[ht]
\centering
\caption{Performance Comparison among Different Schemes}
\label{Table_Compare}
\begin{tabular}{|l|l|l|l|}
\hline
\ & \textbf{\ac{RL}} & \textbf{Min Slice Average} & \textbf{Min Slice Average}  \\ 
\ & \textbf{Reward} & {\bf Throughput Satisfy} & {\bf Delay Satisfy} \\ \hline
{\bf BL-Cen} & 0.801 & 0.738 & 0.739 \\ \hline
{\bf BL-Dist} & 0.948 & 0.952 & 0.962 \\ \hline
{\bf BL-Heur}  &   0.891 &  0.903   & 0.902 \\ \hline
{\bf DIRP} & 0.968 & 0.967 & 0.967 \\ \hline
{\bf Gen}  &   0.961 &  0.960   & 0.961 \\ \hline
{\bf Spec}  &   0.972 &  0.968    & 0.968  \\ \hline
{\bf Spec-Instance}  &   0.971 &  0.970   & \textbf{0.974}  \\ \hline
{\bf Spec-Model}  &   0.962 &  0.965    & 0.966  \\ \hline
{\bf TL-DIRP}  &   \textbf{0.973} &  \textbf{0.971}    & 0.971 \\ \hline
\end{tabular}
\end{table}

Fig. \ref{fig:TL_Reward_Compare} illustrates the evolving rewards during the training and testing processes with different \ac{TL} methods. Note that this comparison is based on max-min fairness in all TL methods. Here we aligned the training process with the Spec scheme for comparison. The results are derived from the average of $3$ independent instances of experiments. Spec with complete knowledge transfer leads to higher reward and robustness compared to Spec-Instance and Spec-Model schemes with partial knowledge at the early stage of the training process, while in the latter two schemes, \ac{TL} also helps in terms of convergence rate, compared with Gen. Furthermore, with offline finetuning Spec-Finetune provides better performance with faster convergence and higher reward within the same training time. In most of the \ac{TL} schemes, we observe that each specialist agent improves its performance with local finetuning from a higher starting point, which helps avoid risky action choices during exploration. With Spec-Instance, the agents behave the worst at the beginning of the training but converge fast later. On the other hand, Spec-Model also suffers from a weaker performance at the beginning and takes a longer time to learn. Eventually, Spec-Instance converges to a similar performance as Spec while Spec-Model achieves a slightly worse performance. Our guess is that there is still a substantial difference between the source domain and the target domain. Without transferring sufficient instances in the source domain (instances following similar distribution to the target domain), the initialized general policy cannot quickly adapt to the target task. Moreover, introducing offline finetuning with the transferred instances to the \ac{TL} scheme further improves the performance by providing even faster convergence and more robust training.

\begin{figure}[t]
	\centering
	\includegraphics[width=.45\textwidth]{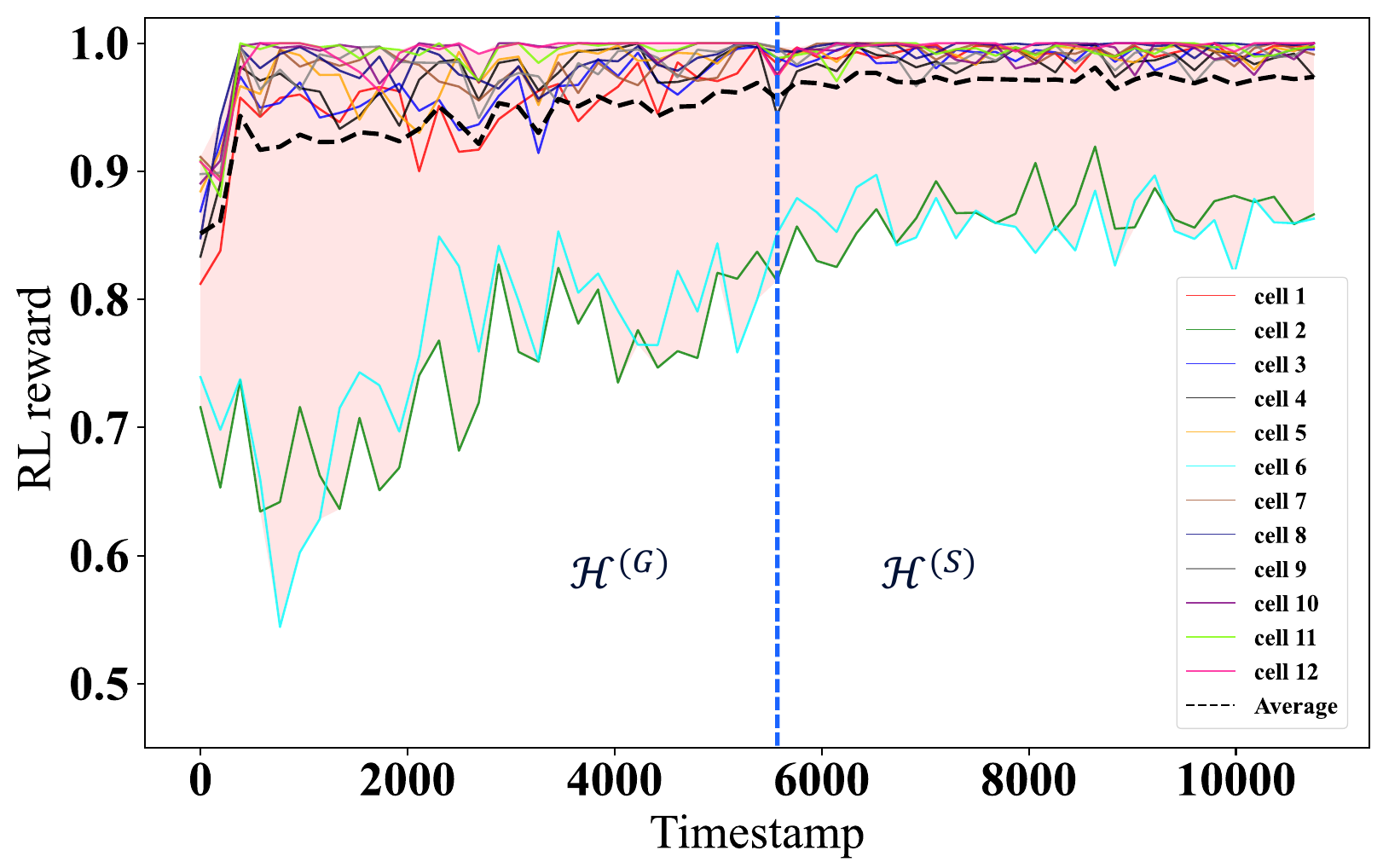}
	\caption{Change of local reward during TL scheme}
	\label{fig:TL_local_reward}
\end{figure}

\begin{figure}[t]
	\centering
	\includegraphics[width=.45\textwidth]{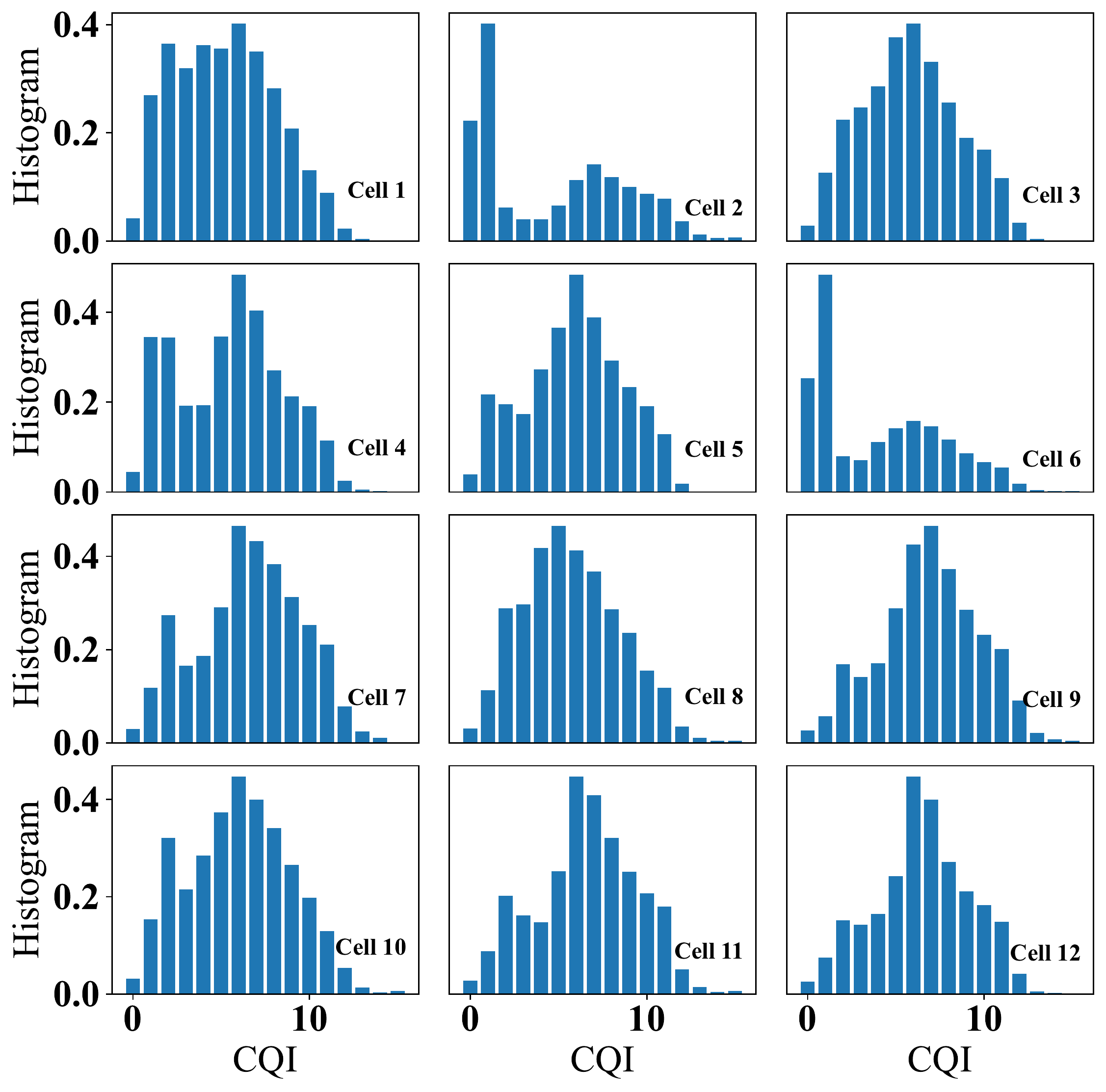}
	\caption{Comparison of \ac{CQI} distribution between cells}
	\label{fig:CQI_Hist}
\end{figure}

In Fig. \ref{fig:TL_local_reward}, we plot the change of local reward in each cell during the complete \ac{TL} procedure from generalist training to specialist finetuning as described in Algorithm \ref{algo:TL-DIRP}. During the time in $\Ho^{\g}$, the local rewards achieved by the generalist agent converge to a generally good reward over all cells. Later, in the local finetuning period $\Ho^{\s}$, the Spec scheme further finetunes the general agent locally and concludes better performance in each cell. The averaged local reward in $\Ho^{\s}$ also indicates better robustness under time-varying traffic demand. 
We can also observe that the rewards from two cells are always lower compared to others during $\Ho^{\g}$, and achieve relatively poor performance after knowledge transfer in $\Ho^{\s}$. Fig. \ref{fig:CQI_Hist} shows the comparison of \ac{CQI} distributions from all cells, it is clear to see that in cell $2$ and cell $6$ which derive poorer performance as shown in Fig. \ref{fig:TL_local_reward} correspondingly, the \ac{CQI} histograms are significantly different compared to others. The difference in user distribution or radio propagation can make the \lq\lq generalist" ambiguous on learning a general policy for all cells, and the derived policy is better for handling the samples from others. Thus, during $\Ho^{\g}$, the rewards in these two cells are lower than others, while in $\Ho^{\s}$, the performances in these two cells get better with local finetuning yet still worse than the others.


Summarized comparisons of the average performance metrics among all schemes in the testing phase are listed in Table \ref{Table_Compare}. We can see that TL-\ac{DIRP} as offline finetuned Spec provides the best performance in terms of the desired \ac{RL} reward and minimum (worst-case) per-slice throughput satisfaction level among the schemes, while Spec-Instance provides a slightly better minimum per-slice delay satisfaction level. Moreover, TL-DIRP encourages a more balanced service quality between all slices in comparison to \ac{TL}-\ac{DIRP}-Log. It is also worth noting that the inference time for a pretrained distributed \ac{DRL} model to make a local decision is less than $4$ milliseconds due to the small sizes of our defined neural networks.

\subsection{Key Takeaways}
In the following, we summarize the takeaways from our numerical analysis:
\begin{itemize}

    \item {\bf Distributed vs. Centralized.} For conventional \ac{DRL} algorithms, the distributed scheme demonstrates good learning capability for adapting to slice-aware traffic and providing good service quality in the defined network scenario with $12$ cells, while the centralized scheme fails to converge to a good reward within the same training time because of its high model complexity and high dimensional state and action spaces. In fact, the larger the scale the network has, the higher the gain the distributed schemes achieve when compared with the centralized approach.
    \item {\bf Inter-agent coordination.} The \ac{DIRP} algorithm with inter-agent coordination and letting the multiple agents share load information provides better performance than the distributed \ac{DRL} in terms of converged reward and convergence rate while maintaining lower model complexity.
    \item {\bf The advantages of transfer learning.} Our proposed \ac{TL}-\ac{DIRP} algorithm further improves the converged reward of \ac{DIRP} with about $11.5\%$ higher start point, $87.5\%$ faster convergence, and lower exploration cost. It is worth noting that, the converged performance of the \ac{TL}-\ac{DIRP} algorithm has higher robustness than \ac{DIRP} without \ac{TL}. It also provides about $15\%$ higher \ac{QoS} satisfaction level for the most critical slice and an $8.8\%$ higher average slice \ac{QoS} fully satisfaction level than the traffic-aware baseline.
    \item {\bf The needs of \lq\lq generalist-to-specialist"~transfer.} During the \lq\lq generalist" training process of \ac{TL}-\ac{DIRP}, the difference in \ac{CQI} between cells make the learning of general policy ambiguous, and the agents from different cell \ac{CQI} derive poorer performance than others. Later in \lq\lq specialist" all agents grant higher reward and robustness with local finetuning.
    \item {\bf Comparison between two reward functions.} The proposed \ac{TL}-\ac{DIRP} approach with reward based on max-min fairness and logarithmic utility can both provide sufficiently good performance among all slice \ac{QoS}. However, max-min fairness reward achieves better \ac{QoS} for critical slice requirements by occupying resources from the slices with less critical requirements, while logarithmic utility provides higher resource efficiency. From the engineering perspective, different reward definitions can be chosen for variant use cases.

    \item {\bf How to transfer.} As for the transferable knowledge, \ac{TL} scheme with combined model and instance transfer enhanced by offline finetuning provides the best performance, in terms of both the starting point and the convergence rate. As expected, when transferring instances only, the local agents still need to train from scratch and suffer from the low performance at the beginning. When transferring the pretrained model, the performance at the beginning is slightly better but requires a longer time to converge. Our guess is that there is substantial difference between the source and target domains according to the \ac{CQI} distribution of cells. Without transferring sufficient instances from the source domain (instances following similar distribution to the target domain), the initialized general policy cannot adapt quickly to the target task.
    Moreover, by introducing an offline finetuning with the transferred instances, \ac{TL}-\ac{DIRP} provides a further performance improvement to the \ac{TL} scheme without offline finetuning in terms of higher start point and faster convergence.
\end{itemize}


\section{Conclusion}\label{sec:concl}

In this paper, we formulated the dynamic inter-cell resource partitioning problem to meet the slice-aware service requirements by jointly optimizing the inter-cell inter-slice resource partitioning. First, we proposed the \ac{DIRP} algorithm to solve the problem with inter-agent coordination. To further improve the algorithm transferability, we designed the \ac{TL}-\ac{DIRP} algorithm by introducing a generalist-to-specialist \ac{TL} framework with different types of transferable knowledge. We evaluated the proposed solutions with a $12$ cells network scenario in a system-level simulator. The evaluation results showed that the \ac{TL}-\ac{DIRP} algorithm provides better slice-aware service performance than the existing baseline approaches. Besides, using \ac{TL} in \ac{MADRL} improves the training performances in different aspects, e.g., higher start point, faster convergence speed, and higher asymptote. We also investigated two reward definitions with max-min fairness and logarithmic utility in \ac{TL}-\ac{DIRP} and found that different rewards should be chosen for variant purposes in practical use cases.


As an extension to the \lq\lq generalist-to-specialist" \ac{TL} scheme, future works include inter-agent \ac{TL}, which enables knowledge transfer from a pretrained \ac{DRL} agent to another, e.g., transferring knowledge from a pretrained cell to a newly deployed cell. However, as we observed in numerical experiments, transferring knowledge between agents with different domains and tasks may deteriorate the performance at the early training phase of \ac{TL}, or, sometimes even cause negative transfer. Thus, quantitative analysis needs to be developed to detect similar \ac{DRL} agents for efficient knowledge transfer.



\bibliographystyle{IEEEtran}
\bibliography{myreferences}

\begin{thebibliography}{10}
\providecommand{\url}[1]{#1}
\csname url@samestyle\endcsname
\providecommand{\newblock}{\relax}
\providecommand{\bibinfo}[2]{#2}
\providecommand{\BIBentrySTDinterwordspacing}{\spaceskip=0pt\relax}
\providecommand{\BIBentryALTinterwordstretchfactor}{4}
\providecommand{\BIBentryALTinterwordspacing}{\spaceskip=\fontdimen2\font plus
\BIBentryALTinterwordstretchfactor\fontdimen3\font minus
  \fontdimen4\font\relax}
\providecommand{\BIBforeignlanguage}[2]{{%
\expandafter\ifx\csname l@#1\endcsname\relax
\typeout{** WARNING: IEEEtran.bst: No hyphenation pattern has been}%
\typeout{** loaded for the language `#1'. Using the pattern for}%
\typeout{** the default language instead.}%
\else
\language=\csname l@#1\endcsname
\fi
#2}}
\providecommand{\BIBdecl}{\relax}
\BIBdecl

\bibitem{Hu2022InterCellReDRL}
T.~Hu, Q.~Liao, Q.~Liu, D.~Wellington, and G.~Carle, ``Inter-cell slicing
  resource partitioning via coordinated multi-agent deep reinforcement
  learning,'' in \emph{IEEE International Conference Communications (ICC)},
  2022.

\bibitem{ksentini2017toward}
A.~Ksentini and N.~Nikaein, ``Toward enforcing network slicing on {RAN}:
  {F}lexibility and resources abstraction,'' \emph{IEEE Communications
  Magazine}, vol.~55, no.~6, pp. 102--108, 2017.

\bibitem{vo2018slicing}
P.~L. Vo, M.~N. Nguyen, T.~A. Le, and N.~H. Tran, ``Slicing the edge:
  {R}esource allocation for {RAN} network slicing,'' \emph{IEEE Wireless
  Communications Letters}, vol.~7, no.~6, pp. 970--973, 2018.

\bibitem{Addad2020OptimizationMF}
R.~A. Addad, M.~Bagaa, T.~Taleb, D.~Dutra, and H.~Flinck, ``Optimization model
  for cross-domain network slices in {5G} networks,'' \emph{IEEE Transactions
  on Mobile Computing}, vol.~19, pp. 1156--1169, 2020.

\bibitem{Beshley2021QoSAwareOR}
H.~Beshley, M.~Beshley, M.~Medvetskyi, and J.~Pyrih, ``{QoS}-aware optimal
  radio resource allocation method for machine-type communications in {5G LTE}
  and beyond cellular networks,'' \emph{Wirel. Commun. Mob. Comput.}, vol.
  2021, pp. 9\,966\,366:1--9\,966\,366:18, 2021.

\bibitem{Fossati2020MultiResourceAF}
F.~Fossati, S.~Moretti, P.~Perny, and S.~Secci, ``Multi-resource allocation for
  network slicing,'' \emph{IEEE/ACM Transactions on Networking}, vol.~28, pp.
  1311--1324, 2020.

\bibitem{Ma2020SlicingRA}
T.~Ma, Y.~Zhang, F.~Wang, D.~Wang, and D.~Guo, ``Slicing resource allocation
  for {eMBB} and {URLLC} in {5G RAN},'' \emph{Wirel. Commun. Mob. Comput.},
  vol. 2020, pp. 6\,290\,375:1--6\,290\,375:11, 2020.

\bibitem{TS123501}
{3GPP, TS 23.501}, ``{System architecture for the 5G System (5GS), V17.4.0},''
  3GPP, March 2022.

\bibitem{Mao2016ResourceMW}
H.~Mao, M.~Alizadeh, I.~Menache, and S.~Kandula, ``Resource management with
  deep reinforcement learning,'' \emph{Proceedings of the 15th ACM Workshop on
  Hot Topics in Networks}, 2016.

\bibitem{ConstrainedRLNetSlicing}
Y.~Liu, J.~Ding, and X.~Liu, ``A constrained reinforcement learning based
  approach for network slicing,'' in \emph{IEEE 28th International Conference
  on Network Protocols (ICNP)}, 2020, pp. 1--6.

\bibitem{DeepSlicingQiang}
Q.~Liu, T.~Han, N.~Zhang, and Y.~Wang, ``{DeepSlicing}: Deep reinforcement
  learning assisted resource allocation for network slicing,'' in \emph{IEEE
  Global Communications Conference (GLOBECOM)}, 2020, pp. 1--6.

\bibitem{Li2018DeepRL}
R.~Li, Z.~Zhao, Q.~Sun, C.-L. I, C.~Yang, X.~Chen, M.~Zhao, and H.~Zhang,
  ``Deep reinforcement learning for resource management in network slicing,''
  \emph{IEEE Access}, vol.~6, pp. 74\,429--74\,441, 2018.

\bibitem{Alqerm2016ACO}
I.~Alqerm and B.~Shihada, ``A cooperative online learning scheme for resource
  allocation in {5G} systems,'' \emph{2016 IEEE International Conference on
  Communications (ICC)}, pp. 1--7, 2016.

\bibitem{Zhao2019DeepRL}
N.~Zhao, Y.-C. Liang, D.~T. Niyato, Y.~Pei, M.~Wu, and Y.~Jiang, ``Deep
  reinforcement learning for user association and resource allocation in
  heterogeneous cellular networks,'' \emph{IEEE Transactions on Wireless
  Communications}, vol.~18, pp. 5141--5152, 2019.

\bibitem{Shao2021GraphAN}
Y.~Shao, R.~Li, Z.~Zhao, and H.~Zhang, ``Graph attention network-based {DRL}
  for network slicing management in dense cellular networks,'' \emph{2021 IEEE
  Wireless Communications and Networking Conference (WCNC)}, pp. 1--6, 2021.

\bibitem{Nie2021MultiAgentDR}
H.~Nie, S.~Li, and Y.~Liu, ``Multi-agent deep reinforcement learning for
  resource allocation in the multi-objective {HetNet},'' \emph{2021
  International Wireless Communications and Mobile Computing (IWCMC)}, pp.
  116--121, 2021.

\bibitem{pan2009survey}
S.~J. Pan and Q.~Yang, ``A survey on transfer learning,'' \emph{IEEE
  Transactions on knowledge and data engineering}, vol.~22, no.~10, pp.
  1345--1359, 2009.

\bibitem{nguyen2021transfer}
C.~T. Nguyen, N.~Van~Huynh, N.~H. Chu, Y.~M. Saputra, D.~T. Hoang, D.~N.
  Nguyen, Q.-V. Pham, D.~Niyato, E.~Dutkiewicz, and W.-J. Hwang, ``Transfer
  learning for future wireless networks: {A} comprehensive survey,''
  \emph{arXiv preprint arXiv:2102.07572}, 2021.

\bibitem{wang2021transfer}
M.~Wang, Y.~Lin, Q.~Tian, and G.~Si, ``Transfer learning promotes 6g wireless
  communications: {R}ecent advances and future challenges,'' \emph{IEEE
  Transactions on Reliability}, 2021.

\bibitem{Parera20}
C.~Parera, Q.~Liao, I.~Malanchini, C.~Tatino, A.~E.~C. Redondi, and M.~Cesana,
  ``Transfer learning for tilt-dependent radio map prediction,'' \emph{IEEE
  Transactions on Cognitive Communications and Networking}, vol.~6, no.~2, pp.
  829--843, 2020.

\bibitem{Taylor2007TransferLV}
M.~E. Taylor, P.~Stone, and Y.~Liu, ``Transfer learning via inter-task mappings
  for temporal difference learning,'' \emph{J. Mach. Learn. Res.}, vol.~8, pp.
  2125--2167, 2007.

\bibitem{zhuang2020comprehensive}
F.~Zhuang, Z.~Qi, K.~Duan, D.~Xi, Y.~Zhu, H.~Zhu, H.~Xiong, and Q.~He, ``A
  comprehensive survey on transfer learning,'' \emph{Proceedings of the IEEE},
  vol. 109, no.~1, pp. 43--76, 2020.

\bibitem{Taylor2009TransferLF}
M.~E. Taylor and P.~Stone, ``Transfer learning for reinforcement learning
  domains: {A} survey,'' \emph{J. Mach. Learn. Res.}, vol.~10, pp. 1633--1685,
  2009.

\bibitem{zhu2020transfer}
Z.~Zhu, K.~Lin, and J.~Zhou, ``Transfer learning in deep reinforcement
  learning: {A} survey,'' \emph{arXiv preprint arXiv:2009.07888}, 2020.

\bibitem{nagib2021transfer}
A.~M. Nagib, H.~Abou-Zeid, and H.~S. Hassanein, ``Transfer learning-based
  accelerated deep reinforcement learning for {5G RAN} slicing,'' in \emph{2021
  IEEE 46th Conference on Local Computer Networks (LCN)}.\hskip 1em plus 0.5em
  minus 0.4em\relax IEEE, 2021, pp. 249--256.

\bibitem{mai2021transfer}
T.~Mai, H.~Yao, N.~Zhang, W.~He, D.~Guo, and M.~Guizani, ``Transfer
  reinforcement learning aided distributed network slicing resource
  optimization in industrial {IoT},'' \emph{IEEE Transactions on Industrial
  Informatics}, 2021.

\bibitem{Zafar2021TransferLI}
H.~Zafar, Z.~Utkovski, M.~Kasparick, and S.~Stańczak, ``Transfer learning in
  multi-agent reinforcement learning with double {Q}-networks for distributed
  resource sharing in {V2X} communication,'' \emph{ArXiv}, vol. abs/2107.06195,
  2021.

\bibitem{TS28530}
{3GPP, TS 28.530}, ``{Technical Specification Group Services and System
  Aspects; Management and orchestration; Concepts, use cases and requirements,
  V17.2.0},'' 3GPP, December 2021.

\bibitem{mo2000fair}
J.~Mo and J.~Walrand, ``Fair end-to-end window-based congestion control,''
  \emph{IEEE/ACM Transactions on Networking}, vol.~8, no.~5, pp. 556--567,
  2000.

\bibitem{bonald2006queueing}
T.~Bonald, L.~Massouli{\'e}, A.~Proutiere, and J.~Virtamo, ``A queueing
  analysis of max-min fairness, proportional fairness and balanced fairness,''
  \emph{Queueing systems}, vol.~53, no.~1, pp. 65--84, 2006.

\bibitem{AutonomicP}
J.~Ewing, ``Autonomic performance optimization with application to
  self-architecting software systems,'' Ph.D. dissertation, 04 2015.

\bibitem{Cavalcante2019ConnectionsBS}
R.~L.~G. Cavalcante, Q.~Liao, and S.~Stańczak, ``Connections between spectral
  properties of asymptotic mappings and solutions to wireless network
  problems,'' \emph{IEEE Transactions on Signal Processing}, vol.~67, pp.
  2747--2760, 2019.

\bibitem{Sciancalepore2018AMI}
V.~Sciancalepore, I.~Filippini, V.~Mancuso, A.~Capone, and A.~Banchs, ``A
  multi-traffic inter-cell interference coordination scheme in dense cellular
  networks,'' \emph{IEEE/ACM Transactions on Networking}, vol.~26, pp.
  2361--2375, 2018.

\bibitem{Xu2018ExperiencedrivenNA}
Z.~Xu, J.~Tang, J.~Meng, W.~Zhang, Y.~Wang, C.~Liu, and D.~Yang,
  ``Experience-driven networking: {A} deep reinforcement learning based
  approach,'' \emph{IEEE INFOCOM 2018 - IEEE Conference on Computer
  Communications}, pp. 1871--1879, 2018.

\bibitem{Song2021ADR}
H.~Song, L.~Liu, J.~D. Ashdown, and Y.~C. Yi, ``A deep reinforcement learning
  framework for spectrum management in dynamic spectrum access,'' \emph{IEEE
  Internet of Things Journal}, vol.~8, pp. 11\,208--11\,218, 2021.

\bibitem{Peng2020DeepRL}
H.~Peng and X.~S. Shen, ``Deep reinforcement learning based resource management
  for multi-access edge computing in vehicular networks,'' \emph{IEEE
  Transactions on Network Science and Engineering}, vol.~7, pp. 2416--2428,
  2020.

\bibitem{Xu2021AggregationTL}
D.~Xu, P.~Qiao, and Y.~Dou, ``Aggregation transfer learning for multi-agent
  reinforcement learning,'' \emph{2021 2nd International Conference on Big Data
  \& Artificial Intelligence \& Software Engineering (ICBASE)}, pp. 547--551,
  2021.

\bibitem{Foerster2016LearningTC}
J.~N. Foerster, Y.~M. Assael, N.~de~Freitas, and S.~Whiteson, ``Learning to
  communicate with deep multi-agent reinforcement learning,'' in \emph{NIPS},
  2016.

\bibitem{Konda1999ActorCriticA}
V.~Konda and J.~Tsitsiklis, ``{Actor-Critic} algorithms,'' in \emph{NIPS},
  1999.

\bibitem{Fujimoto2018AddressingFA}
S.~Fujimoto, H.~V. Hoof, and D.~Meger, ``Addressing function approximation
  error in {Actor-Critic} methods,'' \emph{ArXiv}, vol. abs/1802.09477, 2018.

\bibitem{Silver2014DeterministicPG}
D.~Silver, G.~Lever, N.~Heess, T.~Degris, D.~Wierstra, and M.~A. Riedmiller,
  ``Deterministic policy gradient algorithms,'' in \emph{ICML}, 2014.

\bibitem{kang2019contrastive}
G.~Kang, L.~Jiang, Y.~Yang, and A.~G. Hauptmann, ``Contrastive adaptation
  network for unsupervised domain adaptation,'' in \emph{Proceedings of the
  IEEE/CVF Conference on Computer Vision and Pattern Recognition}, 2019, pp.
  4893--4902.

\bibitem{SeasonII}
N.~S. Networks, ``White paper: Self-organizing network ({SON}): Introducing the
  nokia siemens networks {SON} suite-an efficient, future-proof platform for
  {SON}.'' October, 2009.

\bibitem{Winner}
J.~Meinilä, P.~Kyösti, L.~Hentilä, T.~Jämsä, E.~Suikkanen, E.~Kunnari, and
  M.~Narandžić, ``Wireless world initiative new radio - {Winner+},'' 2010.

\end{thebibliography}

\end{document}